\documentclass[]{aa}  

\usepackage{graphicx}
\usepackage{txfonts}
\usepackage{multirow} 
\usepackage[outdir=./figuras/enpdf/]{epstopdf}
\usepackage{diagbox} 
\usepackage{setspace}
\usepackage{float} 

\usepackage{blindtext}
\usepackage{gensymb}
\usepackage{multirow}
\usepackage[font=small,labelfont=bf]{caption}

\usepackage{enumitem}

\usepackage{caption}
\usepackage{subcaption}

 \usepackage{tabularx}             
    \usepackage{array}           
\usepackage{ amssymb }

\usepackage{natbib,twoopt}
\usepackage[colorlinks,allcolors=blue]{hyperref} 
\usepackage{verbatim}

\usepackage{lscape}

\usepackage{ mathrsfs }

\usepackage{numprint}

\begin{document} 

\newcommand{\doce}{$^{12}$CO }
\newcommand{\docep}{$^{12}$CO}
\newcommand{\trece}{$^{13}$CO }
\newcommand{\trecep}{$^{13}$CO}
\newcommand{\dosuno}{\mbox{$J=$ 2\,--\,1}}
\newcommand{\unocero}{\mbox{$J=$ 1\,--\,0}}
\newcommand{\tresdos}{$J=3-2$}

\newcommand{\seiscincocc}{$J$\,=\,6\,(5)\,$-$\,5\,(4) }
\newcommand{\seiscincoccp}{$J$\,=\,6\,(5)\,$-$\,5\,(4)}
\newcommand{\dosunouu}{$J$\,=\,2\,(1)\,$-$\,2\,(1) }
\newcommand{\dosunouup}{$J$\,=\,2\,(1)\,$-$\,2\,(1)}
\newcommand{\seiscincoud}{$J$\,=\,6\,(1,\,6)\,$-$\,5\,(2,\,3) }
\newcommand{\seiscincoudp}{$J$\,=\,6\,(1,\,6)\,$-$\,5\,(2,\,3)}

\newcommand{\cincocuatro}{$J=5-4$}
\newcommand{\cuatrotres}{$J=4-3$}
\newcommand{\dieznueve}{$J=10-9$}
\newcommand{\ochosiete}{$J=8-7$}
\newcommand{\snso}{$J=68-68$}

\newcommand{\tragua}{$J_{Ka,\,Kc}=6_{1,\,6}-5_{2,\,3}$}
\newcommand{\trsou}{$J_{N}=2_{2}-1_{1}$}
\newcommand{\trsod}{$J=20\,(20)-19\,(20)$}
\newcommand{\trsot}{$J=22\,(22)-21\,(22)$}
\newcommand{\trsoc}{$J_{N}=6_{5}-5_{4}$}
\newcommand{\trsodos}{$J_{Ka,\,Kc}=4_{2,\,2}-4_{1,\,3}$}

\newcommand{\cdsop}{C$^{17}$O}
\newcommand{\cdoop}{C$^{18}$O}
\newcommand{\aguap}{H$_{2}$O}
\newcommand{\vsio}{$^{29}$SiO}
\newcommand{\tsio}{$^{30}$SiO}
\newcommand{\hctn}{HC$_{3}$N}
\newcommand{\htcn}{H$^{13}$CN}
\newcommand{\hcom}{HCO$^{+}$}
\newcommand{\sod}{SO$_{2}$}
\newcommand{\agua}{H$_{2}$O}

\newcommand{\vcero}{$v=0$}
\newcommand{\vuno}{$v=1$}
\newcommand{\vdos}{$v=2$}
\newcommand{\vseis}{$v=6$}

\newcommand{\kms}{\,km\,s$^{-1}$ }
\newcommand{\kmsp}{\,km\,s$^{-1}$}
\newcommand{\ms}{\,M$_{\odot}$ }
\newcommand{\msp}{\,M$_{\odot}$}
\newcommand{\pagb}{post-AGB }
\newcommand{\pagbp}{post-AGB}
\newcommand{\pagbs}{post-AGBs }
\newcommand{\pagbsp}{post-AGBs}

\newcommand{\alls}{AC\,Her, the Red\,Rectangle, 89\,Her, HD\,52961, IRAS\,19157$-$0257, IRAS\,18123+0511, IRAS\,19125+0343, AI\,CMi, IRAS\,20056+1834, and R\,Sct}
\newcommand{\allsp}{AC\,Her, the Red\,Rectangle, 89\,Her, HD\,52961, IRAS\,19157$-$0257, IRAS\,18123+0511, IRAS\,19125+0343, AI\,CMi, IRAS\,20056+1834, and R\,Sct }
\newcommand{\on}{89\,Her }
\newcommand{\onp}{89\,Her}
\newcommand{\iras}{IRAS\,19125+0343 }
\newcommand{\irasp}{IRAS\,19125+0343}
\newcommand{\ac}{AC\,Her }
\newcommand{\acp}{AC\,Her}
\newcommand{\rs}{R\,Sct }
\newcommand{\rsp}{R\,Sct}
\newcommand{\rr}{Red\,Rectangle }
\newcommand{\rrp}{Red\,Rectangle}
\newcommand{\ai}{AI\,CMi }
\newcommand{\aip}{AI\,CMi}

\newcommand{\nir}{NIR excess }
\newcommand{\nirp}{NIR excess}

\newcommand{\fig}{Fig.\,}
\newcommand{\figs}{Figs.\,}
\newcommand{\tab}{Table\,}
\newcommand{\tabs}{Tables\,}

\newcommand{\eq}{Eq.\,}
\newcommand{\eqs}{Eqs.\,}
\newcommand{\sect}{Sect.\,}
\newcommand{\sects}{Sects.\,}
\newcommand{\app}{App.\,}
\newcommand{\secp}{\mbox{\rlap{.}$''$}}

\newcommand{\x}{\,$\times$\,}
\newcommand{\xd}[1]{10$^{#1}$}

\newcommand{\mm}{\,$\pm$\,}

\newcommand{\me}{\textless\,}

\newcommand{\lsim}{\raisebox{-.4ex}{$\stackrel{\sf <}{\scriptstyle\sf \sim}$}}
\newcommand{\gsim}{\raisebox{-.4ex}{$\stackrel{\sf >}{\scriptstyle\sf \sim}$}}

\newcommand\sz{1}

   \title{CO observations in rotating circumbinary post-AGB disks
   }
   \author{I. Gallardo Cava\,\inst{1}, J. Alcolea\,\inst{1}, H. Van Winckel\,\inst{2}, V. Bujarrabal\,\inst{3},
   M. Santander-García\,\inst{1}, and M. Gómez-Garrido\,\inst{1,4}
   }     
    
    \institute{Observatorio Astronómico Nacional (OAN-IGN), Alfonso XII 3, 28014, Madrid, Spain\\                 \email{i.gallardocava@oan.es}
    \and
    Instituut voor Sterrenkunde, KU Leuven, Celestijnenlaan 200B, 3001, Leuven, Belgium
    \and
    Observatorio Astronómico Nacional (OAN-IGN), Apartado 112, 28803, Alcalá de Henares, Madrid, Spain
	\and
	Centro de Desarrollos Tecnológicos, Observatorio de Yebes (IGN), 19141, Yebes, Guadalajara, Spain
	}

\titlerunning{CO observations in rotating circumbinary post-AGB disks}
\authorrunning{Gallardo Cava, I. et al.}

   \date{}
   \date{Received 16 July 2025 / Accepted 28 October 2025}

 
  \abstract
   {
   There is a group of post-asymptotic giant branch (post-AGB) stars that are part of a binary system and that show a significant infrared excess. It is by now well established that these systems have disks around the stellar system with either Keplerian or quasi-Keplerian dynamics, and they have outflows consisting of gas escaping from the rotating disk. These binary post-AGB stars can be categorized into two subclasses depending on the predominance of specific kinematic components: disk-dominated and outflow-dominated sources.
   }
   {
   We present the survey of such sources observed in CO using the IRAM-30\,m telescope, in which we aim to identify the molecular gas in these circumbinary-disk-containing post-AGB nebulae. We aim to analyze the mass distribution of the disk objects studied in CO.
   }
   {
   We present high-sensitivity millimeter-wave observations of the CO line emission from ten binary post-AGB stars. Using the derived formulation and observational data, we calculated the mass of the total gas in the CO-emitting regions of these nebulae. The logarithmic distribution of nebular masses was analyzed using a normal model that incorporates censored data, enabling a more comprehensive analysis and yielding more accurate and representative results.
      }
   {
   CO emission is detected in six post-AGB nebulae from our sample. 
   We have significantly increased the sample of observed sources in CO lines. 
   Some of these objects exhibit very weak molecular emission. Within the sample of disk-containing sources, the total gas mass spans a range of \xd{-4} to \xd{-1}\msp.
   The results derived from this work, along with those from a previous CO single-dish survey and interferometric data, show that this class of binary post-AGB stars exhibits a large range of nebular masses (including both the rotating disk and the expanding component), with variations exceeding a factor of 600. The typical nebular mass of these objects, accounting for both detections and non-detections through censored data analysis, is approximately 2.0\x\xd{-3}\msp.
   }
   {}

   

   \maketitle
%

\section{Introduction}
\label{sec:introduccion}

Previous observations have shown that the circumstellar envelopes (CSEs) around asymptotic giant branch (AGB) stars remain largely spherical on large scales, despite showing asymmetries, and typically expand isotropically at low velocities \citep[10\,$-$\,20\,km\,s$^{-1}$, see][]{sahai2007, castrocarrizo2010, decin2020, danilovich2025}.
Their sucessors, the pre-planetary and planetary nebulae (pPNe and PNe), present different morphologies: pPNe tend to present strongly aspherical and (often) axisymmetrical shapes resulting from the interaction of axial outflows (30\,$-$\,100\,km\,s$^{-1}$) with the CSE formed in the AGB stage; PNe often exhibit shapes that are nearly spherical, axisymmetric, or irregular \citep[see e.g.][]{ueta2000,sahai2007,stanghellini2016}.
This transition lasts only a few thousand years  \citep[][]{balickfrank2002}.

An oft-acknowledged explanation for this evolution is that a companion, either a main-sequence star or a degenerate compact object, accumulates material from a rotating disk. Subsequently, extremely fast jets are propelled, akin to the process observed in protostars \citep[][]{soker2002,frankblackman2004,blackmanlucchini2014}.
Our understanding of the impact of binary systems on post-AGB stars remains limited. Nevertheless, the axisymmetric outflows observed in pPNe and PNe are statistically consistent with the prevalence of binaries in the low- to intermediate-mass star population \citep[][]{demarco2013,demarcoizzard2017}.

Particularly remarkable, certain post-AGB stars, specifically those in binary systems (binary post-AGB stars), consistently exhibit signs of having disks that orbit their central stars. All of these sources show a notable excess in near-infrared (NIR) and narrow CO line profiles indicative of rotating disks \citep[see][and references therein]{vanwinckel2003,bujarrabal2013a}. 
Analysis of their spectral energy distributions (SEDs) indicates the existence of warm dust in proximity to the stellar system, and interferometric infrared data confirm its disk-like structure \citep[][]{hillen2017,corporaal2021,kluska2022}.
These disks are considered stable structures, as is indicated by observations showing significant dust processing, including a high degree of crystallization and the presence of large grains \citep[see e.g.][]{gielen2011a,gielen2011b}.
Additionally, single-dish CO-line observations affirm the widespread occurrence of rotation \citep[][]{bujarrabal2013a}, as the characteristics of the line profiles resemble those found in young stars encircled by a rotating disk \citep[][]{guilloteau2013}. Both post-AGB disks and disks around young stars may look similar, but they present significant
differences in practically all physical-chemical parameters, such as origin, lifetime, chemical abundances, temperature, density, mass, or even size \citep[][]{gallardocava_2023tesis}.

Although the group is very homogeneous, a close inspection of their luminosities has revealed the presence of dusty post-RGB primaries in the sample, as they lie below the canonical AGB lower luminosity limit of about \numprint{2500}\,L$_{\odot}$  \citep{kamath2016}. However, it must be noted that these luminosities, derived through SED fitting \citep{kluska2022}, sometimes present substantial uncertainties due to errors in the distance and reddening corrections \citep[see][]{vanwinckel2025}, so some sources could in fact lie above or below this threshold. In view of this and the structural similarities, we retain the term post-AGB binaries for all dusty evolved binaries discussed in this paper.

The detailed study of these circumbinary post-AGB disks requires interferometric observations \citep[at both radio and optical/IR wavelengths, see e.g.][]{bujarrabal2018, kluska2019}. According to millimeter (mm)-wave interferometric maps, we know that there is an additional extended structure surrounding the rotating disk, which presents low expansion velocities. These outflows may be composed of gas and dust extracted from the disk \citep[see discussion in][]{bujarrabal2016}. 
In this work, we refer to the nebula as all the circumstellar material, including both the well-defined rotating structure that forms the disk and the expanding component.
Recent studies classify these post-AGB binaries into two subclasses: disk-dominated and outflow-dominated, depending on which structure holds most of the total nebular material.
In the disk-dominated subclass, the disk holds the majority of the nebular mass, ranging from 85\% to 95\% of the total nebular material.
In the case of outflow-dominated sources, it is the extended and expanding component that holds the majority of the detected nebular material: 65\% to 75\% \citep[see][]{gallardocava2021,gallardocava2023}.

\begin{table}[t]
 \caption{New circumbinary-disk-containing post-AGB nebulae observed in CO mm-wave lines  through single-dish observations.}
\small
\vspace{-5mm}
\begin{center}

\begin{tabular}{@{}l l c c c@{}}

\hline \hline
 \\[-2ex]
\multicolumn{2}{c}{Binary post-AGB source}   & \multicolumn{2}{c}{Observed coordinates}     & $d$ \\ 
  Name & IRAS      &   \multicolumn{2}{c}{J2000}    & [pc]   \\ \hline
\\[-2ex]

HD\,213985          & 22327$-$1731  & 03:29:07.64  &  +48:18:10.4     & 702$^{+730}_{-680}$  \\[0.5em]
TW\,Cam             & 04166+5719    & 04:20:47.64  &  +57:26:28.5     & 1757$^{+1851}_{-1642}$ \\[0.5em]
RV\,Tau             & 04440+2605    & 04:47:06.73  &  +26:10:45.5     & 1245$^{+1315}_{-1189}$ \\[0.5em]
AY\,Lep             & 05208$-$2035  & 05:22:59.42  &  $-$20:32:53. 0  & 1403$^{+1462}_{-1350}$ \\[0.5em]
HD\,52961           & 07008+1050    & 07:03:39.63  &  +10:46:13.1     & 1813$^{+1923}_{-1707}$ \\[0.5em]
U\,Mon              & 07284$-$0940  & 07:30:47.47  &  $-$09:46:36.8   & 800$^{+916}_{-712}$  \\[0.5em]
HR\,4049            & 10158$-$2844  & 10:18:07.59  &  $-$28:59:31.0   & 1449$^{+1759}_{-1263}$  \\[0.5em]
AF\,Crt             & 11472$-$0800  & 11:49:48.00  &  $-$08:17:20.4   & 6107$^{+9132}_{-4134}$ \\[0.5em]
V\,Vul              & 20343+2625    & 20:36:32.02  &  +26:36:14.5     & 1083$^{+1144}_{-1041}$ \\[0.5em]
AU\,Peg             & 21216+1803    & 21:24:00.24  &  +18:16:43.8     & 597$^{+603}_{-591}$                                         
\\[0.5em]

\hline
 
\end{tabular}

\end{center}
\small
\vspace{-1mm}
\textbf{Notes.} Distances, $d$, and their associated upper and lower uncertainties are adopted from \citet{bailerjones2021}.
\label{tab:fuentes_co}
\end{table}

A remarkable increase in the number of nebulae around post-AGB binaries studied in CO through mm-CO single-dish observations is reported in this work. This is important, as these objects have not been widely studied using single-dish observations, except for the previous CO survey \citep[][]{bujarrabal2013a}, or the more recent survey searching for other species by \cite{gallardocava2022}, which focused on molecular content.

Motivated by these objectives, this paper is laid out as follows. 
We describe the observational techniques used in \sect\ref{sec:observations}. We introduce the observed sample in \sect\ref{sec:final_co_data}. In this section, we also provide the main parameters derived from the final CO spectra. In \sect\ref{sec:calculo_masas_co}, we calculate the nebular masses of these disk-containing post-AGB nebulae. 
Next, in \sect\ref{sec:comparativa}, we compare the nebular mass values obtained in this work with those from earlier studies. In \sect\ref{sec:othermolecules}, we discuss the presence of molecules other than CO.
Lastly, we provide a concise overview of our results in \sect\ref{sec:conclusiones}.

\begin{figure}[t]
\center
\includegraphics[width=\sz\linewidth]{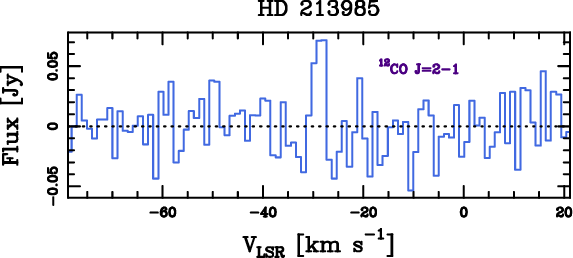} 
\caption{\small Newly detected \doce\dosuno\ line profile in HD\,213985. The x axis indicates velocity with respect to $V_{\text{LSR}}$ and the y axis represents the detected flux measured in janskys.}
    \label{fig:213985_co}  
\end{figure}

\begin{figure}[t]
	\center
	\includegraphics[width=\sz\linewidth]{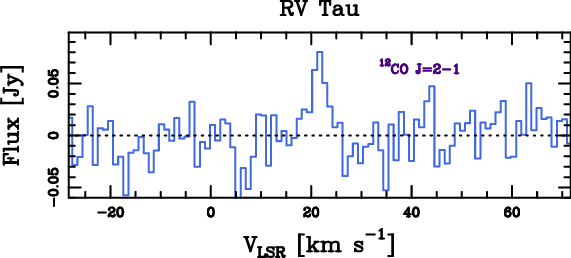} 
	\caption{\small Newly detected \doce\dosuno\ line profile in RV\,Tau. 
		Axes as in \fig\ref{fig:213985_co}.
	}
	\label{fig:rvtau_co}  
\end{figure}

\section{Observations and data reduction}
\label{sec:observations}

In this work, we observed a selected sample of disk-containing post-AGB nebulae through single-dish observations. The CO lines were studied using the IRAM-30\,m telescope, while the SiO maser line was investigated using the Yebes-40\,m telescope.

\subsection{IRAM-30\,m data}

Our data for the \doce and \trece \dosuno\ and \unocero\ lines have been taken using the IRAM-30\,m telescope
using the Eight MIxer Receiver (EMIR), which can simultaneously observes in the 1.3 and 2.6\,mm bands in dual polarization mode \citep[EMIR,][]{carter2012}.
These observations were performed in two different runs under project 144-17 for 125\,hours in total. Data from the first run were obtained between 31 January and 5 February 2018, while data from the second one were obtained between 10 May and 14 May 2018.

We connected the Fast Fourier Transform Spectrometer (FFTS) units to the EMIR receiver with a resolution of 200\,kHz per channel, which corresponds to 0.25 and 0.5\,km\,s$^{-1}$ in the 1.3\,mm and 2.6\,mm bands, respectively.
The spatial resolution, or half power beam width (HPBW), of the observations is 
$\sim$\,22$''$ and $\sim$\,11$''$ at the frequencies of CO \unocero\ and \dosuno\ lines, respectively.
Our sources were observed using the wobbler-switching mode, which provides flat baselines. The subreflector was alternatively placed between ON and OFF positions every 2\,seconds, with the OFF reference being symmetrically located at $\pm$\,90$''$ in the azimuth direction. 
We simultaneously observed using the vertically and horizontally polarized receivers, with no significant variations in their relative calibrations.

Pointing and focus were checked every four hours, with particular attention during the temperature variations associated with dusk and dawn. All corrections were negligible compared to the beam size.
The flux calibration was derived using the chopper-wheel method, repeated at $\sim$15-minute intervals. Supplementary corrections were obtained from observations of the standard calibrators NGC\,7027 and CW\,Leo (IRC+10216).

\subsection{Yebes-40\,m data}

We conducted observations in the 7\,mm band (\textit{Q} band) using the Yebes-40\,m telescope. Five sources from our sample were observed between 20 November 2024 and 13 January 2025, with a total on-source time of 50 hours under project code 24B011: HD\,213985, RV\,Tau, U\,Mon, HR\,4049, and IRAS\,11472$-$0800. \textit{Q}-band data for HD\,52961 were obtained from project 20B006 and are already published. Detailed information on these sources is found in the next section.

The received signal was processed using the FFTS backend units. The \textit{Q} band provided an instantaneous bandwidth of 18\,GHz, a spectral resolution of 38\,kHz, and an HPBW ranging from 37$''$ to 49$''$. Observations were carried out using the wobbler-switching technique, with a typical offset of aroud 300$''$ in the azimuthal direction, resulting in flat baselines. Spectra were simultaneously obtained in both vertical and horizontal linear polarizations, with no significant differences found in their flux density calibrations. For additional technical details, see \citet{tercerof2021}.

\subsection{Data reduction}

The data reduction process was carried out with CLASS within the GILDAS\,\footnote{GILDAS is a software package focused in reducing and analyzing mainly mm observations from single-dish and  interferometric telescopes. It is developed and maintained by IRAM, LAOG/Université de Grenoble, LAB/Observatoire de Bordeaux, and LERMA/Observatoire de Paris. Continuum and Line Analysis Single-dish Software (CLASS) is part of the GILDAS software package. For further details, see \url{https://www.iram.fr/IRAMFR/GILDAS}} software package. For each source, we implemented the standard data reduction procedure, involving the exclusion of bad scans, averaging of valid ones (for both polarizations), and the subtraction of a first-order polynomial baseline.
Also, the derived calibration corrections imply that the accuracy of the absolute scale is within a margin of $<$\,20\%.

The observational data obtained from both the IRAM-30\,m and Yebes-40\,m telescopes are expressed in units of $T_{\text{A}}^{*}$, which represents the antenna temperature corrected for atmospheric attenuation, rearward spillover, and ohmic losses. This calibration provides a consistent measure of the received signal, but further conversion is required for physical interpretation. Thus, the data were converted to flux density (Jy) or the main-beam temperature ($T_{\text{mb}}$), depending on the analysis needs, using the telescope-specific efficiency parameters provided by IRAM\,\footnote{\url{https://publicwiki.iram.es/Iram30mEfficiencies}} and Yebes\,\footnote{\url{https://rt40m.oan.es/rt40m_en.php}}.

\begin{figure}[t]
\center
\includegraphics[width=\sz\linewidth]{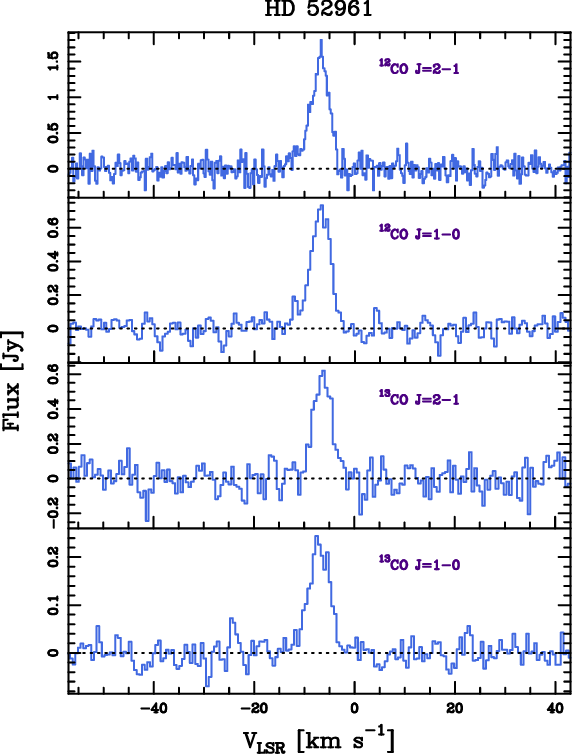} 
\caption{\small Newly detected CO line profiles in HD\,52961. Axes as in \fig\ref{fig:213985_co}.}
    \label{fig:hd52961_co}  
\end{figure}

\begin{figure}[t]
\center
\includegraphics[width=\sz\linewidth]{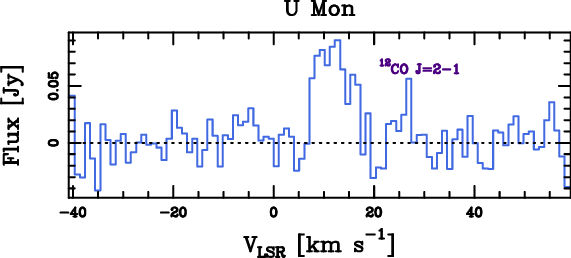} 
\caption{\small Newly detected \doce\dosuno\ line profile in U\,Mon. Axes as in \fig\ref{fig:213985_co}.}
    \label{fig:umon_co}  
\end{figure}

\begin{figure}[t]
\center
\includegraphics[width=\sz\linewidth]{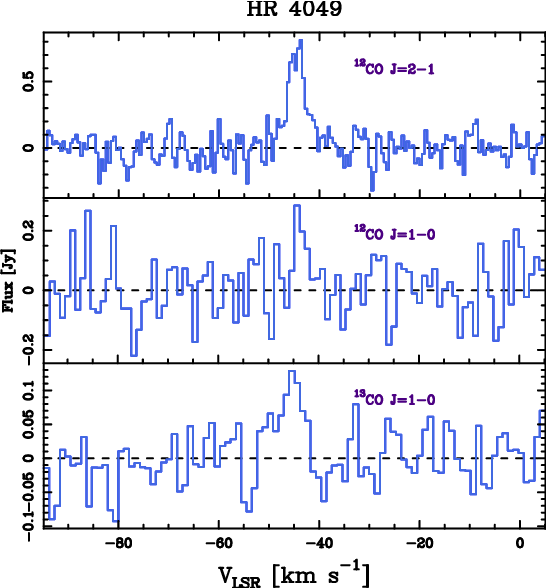} 
\caption{\small Newly detected CO line profiles in HR\,4049. Axes as in \fig\ref{fig:213985_co}.}
    \label{fig:hr4049_co}  
\end{figure}

\begin{figure}[t]
\center
\includegraphics[width=\sz\linewidth]{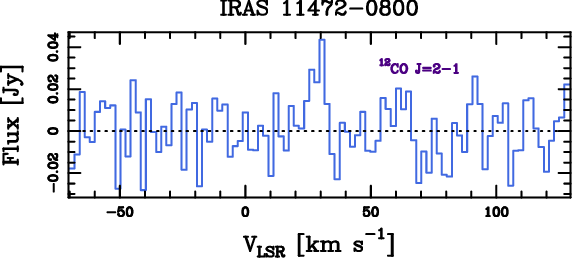} 
\caption{\small Newly detected \doce\dosuno\ line profile in IRAS\,11472$-$0800. Axes as in \fig\ref{fig:213985_co}.
}
    \label{fig:i11472_co}  
\end{figure}

\section{Source sample and CO data}
\label{sec:final_co_data}

Ten sources have been observed in this second single-dish CO survey project: HD\,213985, TW\,Camelopardalis, RV\,Tauri, IRAS\,05208$-$2035, HD\,52961, U\,Monocerotis, HR\,4049, IRAS\,11472$-$0800, V\,Vulpeculae, and AU\,Pegasi (see \tab\ref{tab:fuentes_co}).
This work provides a significant expansion in the number of disk-containing post-AGB sources observed in mm-waves through CO single-dish observations \citep[see][for details of the first single-dish CO survey]{bujarrabal2013a}.

These ten new sources are identified as binary post-AGB stars. Their evolved nature is evidenced by their high luminosities, which range from $\sim$\,70 to $\sim$\,12\,600\,L$_{\odot}$ with a mean value of $\sim$\,2\,700\,L$_{\odot}$ (private communication), and their intermediate spectral types \citep[][]{oomen2019,kluska2019}. Additionally, all of them show significant NIR excess in their SED, together with far-infrared (FIR) excess that is indicative of material ejected by the star.

In \citet{bujarrabal2013a}, 24 such circumbinary-disk-containing post-AGB nebulae were observed in CO lines; 3 of them are also observed in detail in this work. The 7 remaining new sources represent a 30\% increase in the CO sample.
Together, these 31 objects studied in CO lines represent $\sim$\,37\% of the total sample of this type of disk sources, which currently numbers 85.

Nevertheless, we recall that the luminosities derived from SED fitting can sometimes suffer from substantial uncertainties, mainly due to errors in distance estimates and reddening corrections \citep[see][]{kluska2022,vanwinckel2025}. This implies that some of the objects might actually fall below the canonical lower luminosity limit for AGB stars \citep[$\sim$\,2\,500\,L$_\odot$;][]{kamath2016}, in which case they could be classified as post-RGB rather than post-AGB binaries \citep{moltzer2025}. We therefore adopt the generic term post-AGB binaries throughout this paper for consistency.

We delve into the key findings obtained from the CO survey for each of these new sources:

\begin{itemize}

\item HD\,213985: This source exhibits a narrow and faint \mbox{$^{12}$CO $J=2-1$} line profile. This tentative detection is centered at $-$29\kms (see \fig\ref{fig:213985_co}).
Although this velocity differs from that reported by \citet{oomen2018} (see \app\ref{sec:velocidades} for a detailed discussion\,\footnote{\app\ref{sec:velocidades} presents a comparison between the heliocentric velocities derived from CO line observations (including both new and previously published data) and those determined by \citet{oomen2018} from radial velocity curves. A cross-match was performed to identify sources common to both samples, enabling a direct comparison of their velocity estimates.}), the molecular line profile suggests that the emission could still originate from circumstellar material physically linked to this source.

\item TW\,Cam: We find no CO emission for this source. Strong ISM contamination is detected in the range $-$35 to $-$40\kmsp.

\item RV\,Tau: This source was previously observed with no detections. Our new and improved observations benefit from a longer integration time, allowing us to clearly detect a narrow \doce\dosuno\ line (see \fig\ref{fig:rvtau_co}).

\item IRAS\,05208$-$2035: This source does not show any observable CO emission.

\item HD\,52961: This object shows intense \doce\ and \trece\ lines (see \fig\ref{fig:hd52961_co}).
It exhibits a narrow line profile, characteristic of rotating disks, along with relatively wide line wings. The \unocero\ lines (both in \doce\ and \trecep) show the
distinctive double peak commonly seen in rotating disks. From both observational and theoretical perspectives, it is widely accepted that these profiles effectively indicate the presence of rotating disks \citep[][]{bujarrabal2013a, gallardocava2021}. The additional intensity in the line wings appears difficult to account for solely based on disk emission, so the presence of a relatively massive outflow could explain this CO line profile \citep[as in the case of R\,Sct or AI\,CMi, see][]{bujarrabal2013a}.

\item U\,Mon: The first CO survey did not yield any detections for this source. 
The superior quality of these recent observations has facilitated the clear detection of the \doce\dosuno\ line.
We are able to tentatively discern the double peak in this narrow CO line, which does not appear to show line wings (see \fig\ref{fig:umon_co}), in a similar way to that found in AC\,Her \citep[see][]{bujarrabal2013a, gallardocava2021}.

\item HR\,4049: This source was already observed in the first survey, yielding weak CO line detections. Our new observations show significant improvement over the previous study: the signal-to-noise ratios for the \doce\dosuno, \doce\unocero, and \trece\unocero\ transitions are now 8, 2.5, and 3.4, representing improvement factors of 1.5, 1.3, and 1.3, respectively. The resulting higher-quality data, particularly for the \doce\dosuno\ and \trece\unocero\ lines, more useful for our study, lead to a much more accurate outcome (see \fig\ref{fig:hr4049_co}).

\begin{table*}[h]
\caption{Summary of observational results.}
\small
\tiny
\vspace{-5mm}
\begin{center}

\begin{tabular*}{\textwidth}{@{\extracolsep{\fill\quad}}lllcccccccl}
\hline \hline
\noalign{\smallskip}

\multirow{2}{*}{Source} &\multirow{2}{*}{Molecule} &\multirow{2}{*}{Transition}  & $I$\,(peak) & $\sigma$ & $\int I\,\text{d}V$ & $\sigma \left(\int I\,\text{d}V \right)$ & $\delta V$ & $V_{\text{LSR}}$ &  \multirow{2}{*}{Comments} \\
& &  & [Jy] & [Jy] & [Jy\,km\,s$^{-1}$] & [Jy\,km\,s$^{-1}$] & [km\,s$^{-1}$]  & [km\,s$^{-1}$] & \\
\hline
\\[-2ex]

HD\,213985     &   \doce      &   \dosuno         &  7.7E-02    &   3.2E-02    &   1.9E-01    &   8.8E-02  &   1.02    & $-$29      &  Tentative detection           \\
     &                  &   \unocero         &   $\leq$\,1.1E-01    &   3.7E-02    &   $\leq$\,3.0E-01    &   1.0E-01  &    1.02        &             &      \\
      &    \trece              &   \unocero         &  $\leq$\,5.1E-02    &   1.7E-02    &   $\leq$\,1.5E-01    &   4.8E-02    &  1.06     &        &     \\  

\hline    

\\[-2ex] 
 
TW\,Cam     &   \doce      &   \dosuno         &  $\leq$\,7.5E-02    &   2.5E-02    &   $\leq$\,3.4E-01    &   1.1E-01    &   1.02     &               &   ISM contamination           \\
     &                  &   \unocero         &   $\leq$\,1.1E-01    &   3.7E-02    &   $\leq$\,5.1E-01    &   1.7E-01   &     1.02         &             &   ISM contamination          \\
      &    \trece              &   \unocero         &   $\leq$\,6.6E-02    &   2.2E-02    &   $\leq$\,3.1E-01    &   1.0E-01  &  1.06     &            &    ISM contamination          \\       
      
\hline   
   
\\[-2ex]

RV\,Tau     &   \doce      &   \dosuno         &  8.7E-02    &   2.2E-02    &   2.5E-01    &   7.3E-02 &    1.02         &     21.5         &      ISM contamination        \\
     &                  &   \unocero         &  $\leq$\,7.2E-02    &   2.4E-02    &   $\leq$\,2.4E-01    &   8.0E-02        &             1.02 &             &     ISM contamination        \\
      &    \trece              &   \unocero         &  $\leq$\,3.6E-02    &   1.2E-02    &   $\leq$\,1.2E-01    &   4.0E-02   &      1.02 &            &     ISM contamination         \\    
      
\hline     
\\[-2ex]  

IRAS\,05208$-$2035     &   \doce   &   \dosuno         &   $\leq$\,8.1E-02    &   2.7E-02    &   $\leq$\,3.7E-01    &   1.2E-01  &    1.02    &       &      \\
     &                  &   \unocero         &   $\leq$\,1.7E-01    &   5.7E-02    &   $\leq$\,7.8E-01    &   2.6E-01   &   1.02           &             &      \\
      &    \trece              &   \unocero         &   $\leq$\,6.3E-02    &   2.1E-02    &   $\leq$\,3.0E-01    &   1.0E-01  &  1.06     &        &      \\    
     
\hline  
      
\\[-2ex] 

HD\,52961   &   \doce    &   \dosuno         &    1.9E+00    &   1.3E-01    &   7.6E+00    &   2.4E-01   &  0.25   &   $-$7.0      \\
     &                  &   \unocero         &  7.4E-01    &   5.1E-02    &   3.5E+00    &   1.3E-01 &   0.51     &  $-$7.0      &             \\  
      &    \trece              &   \dosuno         & 6.6E-01    &   7.9E-02    &   2.9E+00    &   1.8E-01 &  0.53     &    $-$7.0  &  \\      
      &                  &   \unocero         & 2.4E-01    &   2.4E-02    &   1.2E+00    &   8.0E-02 & 0.53  &  $-$7.0   &              \\      
      
\hline

\\[-2ex]

U\,Mon     &   \doce     &   \dosuno         &    9.5E-02    &   1.9E-02    &   6.1E-01    &   8.2E-02 &   1.02    &     10.5  &     \\
     &                  &   \unocero         &   $\leq$\,6.9E-02    &   2.3E-02    &   $\leq$\,3.0E-01    &   9.9E-02   &             1.02 &             &     \\
      &    \trece              &   \dosuno         &   $\leq$\,1.8E-01    &   6.1E-02    &   $\leq$\,7.9E-01    &   2.6E-01    &      1.02 &     &      \\    
      &                  &   \unocero         &    $\leq$\,2.9E-02    &   9.6E-03    &   $\leq$\,1.2E-01    &   4.1E-02   &      1.02  &     &     \\

\hline  
    
\\[-2ex] 
 
HR\,4049    &   \doce   &   \dosuno         &    8.8E-01    &   1.1E-01    &   3.7E+00    &   2.4E-01  &    0.51   &  $-$45.0        &        \\
     &                  &   \unocero         &  2.8E-01    &   1.1E-01    &   7.1E-01    &   3.2E-01 &  1.02      &   $-$45.0          &     \\  
      &     \trece     &   \unocero  & 1.3E-01    &   3.8E-02    &   5.4E-01    &   1.5E-01  &            1.06 &  $-$45.0      &     \\     

\hline    
 
\\[-2ex] 
 
IRAS\,11472$-$0800     &   \doce       &   \dosuno         &  4.6E-02    &   1.5E-02    &   2.4E-01    &   8.3E-02 &  2.03 &  30.0   &  ISM contamination   \\
     &                  &   \unocero         &   $\leq$\,7.8E-02    &   2.6E-02    &   $\leq$\,2.2E-01    &   7.3E-02       &   1.02  &             &  ISM contamination   \\
      &    \trece              &   \unocero         &  $\leq$\,3.6E-02    &   1.2E-02    &   $\leq$\,1.0E-01    &   3.4E-02    &  1.02     &            &    ISM contamination  \\

\hline  
  
\\[-2ex] 

V\,Vul     &   \doce      &   \dosuno         &    $\leq$\,8.7E-02    &   2.9E-02    &   $\leq$\,4.0E-01    &   1.3E-01   &   1.02      &      &     \\
     &                  &   \unocero         &   $\leq$\,1.1E-01    &   3.7E-02    &   $\leq$\,5.1E-01    &   1.7E-01  &   1.02    &             &             \\
      &    \trece              &   \unocero         &     $\leq$\,5.4E-02    &   1.8E-02    &   $\leq$\,2.6E-01    &   8.5E-02  & 1.06  &        &       \\

\hline  

\\[-2ex] 
  
AU\,Peg     &   \doce       &   \dosuno         &   $\leq$\,1.8E-01    &   5.9E-02    &   $\leq$\,8.1E-01    &   2.7E-01  &    1.02     &               &              \\
     &                  &   \unocero         &   $\leq$\,1.6E-01    &   5.5E-02    &   $\leq$\,7.5E-01    &   2.5E-01 &   1.02      &             &             \\
      &    \trece              &   \unocero         &   $\leq$\,8.1E-02    &   2.7E-02    &   $\leq$\,3.8E-01    &   1.3E-01 &   1.06    &          &     \\

\hline
\end{tabular*}

\end{center}
\small
\vspace{-1mm}
\textbf{Notes.} Sources with upper limits lack CO detections. $V_{\text{LSR}}$ indicates the centroid of the CO line.

\label{tab:lineas_detectadas}
\end{table*}

\item IRAS\,11472$-$0800: This source also shows the narrow CO line profile of rotating disks, where the double peak can be tentatively seen (see \fig\ref{fig:i11472_co}).
Note that the x axis shows a velocity width of 200\kmsp.

\item V\,Vul: There is no detected CO emission from this source.

\item AU\,Peg: No CO emission was detected from this source.

\end{itemize}

As a summary of what has been discussed, we have detected CO emission in HD\,213985, RV\,Tau, U\,Mon, HR\,4049, HD\,52961, and IRAS\,11472$-$0800. Only HD\,52961 presents intense \doce\ and \trece\ lines compared with the rest of the surveyed sources and in line with other intense objects of our first CO survey, such as R\,Sct or the Red\,Rectangle.
All of these CO-emitting sources show a narrow component in their line profiles, suggesting the presence of a rotating structure in their nebulae
(see \figs\ref{fig:213985_co}, \ref{fig:rvtau_co}, \ref{fig:hd52961_co}, \ref{fig:umon_co}, \ref{fig:hr4049_co}, and \ref{fig:i11472_co}).
For the sources without CO emission, we have estimated upper limits.

The main observational parameters are given in \tab\ref{tab:lineas_detectadas}.
The table lists the peak flux, $I$\,[Jy], its associated noise, $\sigma$\,[Jy], the integrated line intensity, $\int\,I\,\text{d}V$\,[Jy\,km\,s$^{-1}$], its associated uncertainty, $\sigma \left(\int\,I\,\text{d}V\right)$\,[Jy\,km\,s$^{-1}$], the spectral resolution, $\delta V$\,[km\,s$^{-1}$], the velocity centroid of the CO line measured with respect to the local standard of rest, $V_{\text{LSR}}$\,[km\,s$^{-1}$], and some relevant comments.

Uncertainty in the integrated line intensity for CO detections was estimated using the following expression:
\begin{equation}
\sigma \left(\int\,I\,\text{d}V\right) = \sigma \ \delta V \sqrt{n_{\text{ch}}} ,
\label{eq:incertidumbre_area}
\end{equation}
where $n_{\text{ch}}$ is the number of channels covered by the detected CO line.
For the undetected lines, we used an upper limit of \mbox{$3\sigma \times \sqrt{n_{\text{ch}}} \times \delta V$}, where $n_\text{ch}$ corresponds to the number of channels typically spanned by the \doce\dosuno\ line. When a CO line is detected, $n_\text{ch}$ is taken as the number of channels covered by that line in the source. In contrast, when no line is detected, $n_\text{ch}$ is taken as the average number of channels measured from all the detected lines in this work.
The number of channels for this line is 52 for HD\,52961, 7 for HD\,213985, 11 for RV\,Tau, 18 for U\,Mon, 19 for HR\,4049, and 8 for IRAS\,11472$-$0800, with an average of 20 channels across all sources.

\section{Total mass of circumbinary-disk-containing post-AGB nebulae}
\label{sec:calculo_masas_co}

This section describes the mathematical formalism adopted for calculating the masses of CO-emitting regions using observational data. Subsequently, we shall discuss and interpret this data to extract physical information from the observed objects.

\subsection{Calculation methodology}
\label{sec:metodologia}

The methodology used in this work to estimate masses of disk-containing nebulae is the same to that used in \citet{bujarrabal2013a,gallardocava2022}. First, the main-beam temperature ($T_{\text{mb}}$) is defined as:
\begin{equation} \label{eq:tmb}
T_{\text{mb}} \, (V_{\text{LSR}}) = S_{\nu} \, \tau \, \phi\,(V_{\text{LSR}}) \, \Omega_{\text{S}} \, \Omega_{\text{mb}}^{-1} ,
\end{equation}
where each term is defined as follows:

\begin{itemize}

\item
$\Omega_{\text{S}}$ is the solid angle subtended by the source, expressed as
\begin{equation}
\Omega_{\text{S}} = 2\pi \, (1 - \cos \alpha) \approx \pi \, \alpha^{2} \approx \pi \, \frac{r^{2}}{d^{2}} \approx \frac{s}{d^{2}},
\end{equation}
where $\alpha$ is the angle formed by the distance, $d$, and the radius, $r$, of the observed source. 
Considering that $d$ will always take large values due to long distances, $\alpha$ will always have small values. Thus, we can approximate \mbox{$\tan \alpha \approx \frac{r}{d}$} and \mbox{$\cos \alpha \approx 1 - \frac{\alpha
^{2}}{2}$}.
The quantity $s$ is the surface area of the observed source derived from its radius and its distance.

\item
$\Omega_{\text{mb}}$ is the solid angle subtended by the telescope main beam:
\begin{equation}
\Omega_{\text{mb}} = \frac{\pi}{4\ln 2} \, \theta_{\text{HPBW}}^{2} \approx 1.133 \, \theta_{\text{HPBW}}^{2}, 
\end{equation}
in which $\theta_{\text{HPBW}}$ is the HPBW.

\item 
$\phi\,(V_{\text{LSR}})$ represents the normalized line profile in terms of Doppler-shifted velocity and it takes a value of 1 when it is integrated in terms of velocity.

\item
$S_{\nu}$ is the source function and it is given by
\begin{equation}
S_{\nu} = \frac{h\nu}{k_{\text{B}}} \left( \frac{1}{\exp{\frac{h\nu}{k_{\text{B}}T_{\text{ex}}}}-1} - \frac{1}{\exp{\frac{h\nu}{k_{\text{B}}T_{\text{BG}}}}-1}
\right)  ,
\end{equation}
where $k_{\text{B}}$ is the Boltzmann constant, $T_{\text{ex}}$ represents the excitation temperature of the molecular transition, and $T_{\text{BG}}$ is the background temperature, with a constant value of 2.73\,K.

\item 
The characteristic optical depth, $\tau$, is
\begin{equation}
\tau = \frac{c^{3}}{8\pi\nu^{3}} \, A_{\text{ul}} \, g_{\text{u}} \, (x_{\text{l}} - x_{\text{u}}) \, n_{\text{T}} \, X \, L   ,
\end{equation}
where $X$ represents the abundance of the molecule relative to the total number of particles (\doce\ and \trece\ in this work), $n_{\text{T}}$ is the total number density of particles, and $L$ represents the typical length of the studied source along the line of sight. Thus, this opacity expression can be rewritten in terms of the mass, $M$, taking into account that
\mbox{$M = n_{\text{T}} \, L \, s = n \, M_{\text{mol}} \, L \,s $},
with $M_{\text{mol}}$ being the total mass of the molecule-rich emitting gas and $n$ the number of particles.
$A_{\text{ul}}$ is the Einstein coefficient of the molecular transition, where $u$ and $l$ are the upper and lower levels of the transition, respectively.
The statistical weight of the upper level ($g_{\text{u}}$) is
\begin{equation}
g_{\text{u}} \equiv g_{J} = 2 \, J + 1.
\end{equation}
Upper and lower level populations can be expressed as
\begin{equation}
x_{J} \sim \frac{\exp\left(-E_{J} T_{\text{rot}}^{-1}\right)}{F(T_{\text{rot}})}     ,
\end{equation}
in which the energy of the level is calculated assuming a simple linear rotor and taking into account the rotational constant of the molecule ($B_{\text{rot}}$ is around 2.8 and 2.6\,K for \doce\ and \trecep, respectively):
\begin{equation}
E_{J} = B_{\text{rot}} \, J \, (J + 1) .
\end{equation}

The rotational temperature ($T_{\text{rot}}$) is the typical temperature value that represents the average of the excitation temperatures of the relevant rotational transitions used, i.e.,\ low-$J$ transitions in our case. This temperature is used to estimate the partition function:
\begin{equation}
F \, (T_{\text{rot}}) = \sum_{J} g_{J} \, \exp \left(-\frac{E_{J}}{T_{\text{rot}}} \right) = \frac{T_{\text{rot}}}{B_{\text{rot}}} .
\end{equation}

\end{itemize}

Finally, integrating \eq\ref{eq:tmb} in terms of velocity, we get
\begin{equation} \label{eq:area_masa}
\begin{split}
\quad \int T_{\text{mb}} \, & \left(V_{\text{LSR}}\right) \, \text{d}V =  \\
&  \hspace{-5mm} 
S_{\nu} \, \frac{c^{3}}{8\pi\nu^{3}} \, A_{\text{ul}} \, g_{\text{u}} \, (x_{\text{l}} - x_{\text{u}}) \, X \, \frac{4\,\ln 2}{\pi} \,\theta_{\text{HPBW}}^{2} \, M_{\text{mol}} \, d^{-2} .
\end{split}
\end{equation}

Therefore, by following this procedure, we obtain an analytical expression that directly relates the area (velocity integrated main beam temperature) of the molecular rotational line to the nebular mass of the CO-emitting region, additionally being inversely proportional to the square of its distance: $\text{Area} \equiv \int T_{\text{mb}} \, (V_{\text{LSR}}) \, \text{d}V \propto M_{\text{mol}} \, d^{-2}$.

The application of this procedure relies on the assumption of optically thin emission, a valid consideration in our case due to the notably weak emission of the molecular lines.
Other factors can influence the opacity, such as the inclination angle of the disk. In systems where the disk is seen edge-on, the CO lines tend to be more optically thick, particularly near the systemic velocity.
As in the previous CO single-dish work, we assume \mbox{$T_{\text{rot}} = 50$\,K}, which is expected for kinetic temperatures between 70 and 100\,K and total densities between approximately 10$^{4}$ and 10$^{5}$\,cm$^{-3}$ \citep[see details in][]{bujarrabal2013a}.
The gas mass values were derived by assuming an average particle mass of \mbox{3\x\xd{-24}\,g} that takes into account molecular hydrogen (H$_2$), atomic hydrogen, and helium. We also assumed a relative abundance of \mbox{$X$($^{12}$CO) =  2\x\xd{-4}}, which may be a low value when compared to AGB CSEs \citep[\mbox{2\,$-$\,4\x\xd{-4}} in O-rich stars and \mbox{6\,$-$\,8\x\xd{-4}} in C-rich stars, see][]{teyssier2006,ramstedt2008}, but is in line with previous works on this type of objects and with the molecular study performed on these sources \citep[see e.g.][]{bujarrabal2013a, gallardocava2021}. 
We also considered a $^{12}$CO/$^{13}$CO abundance ratio of 10, which is also the typical ratio found in these objects \citep[see e.g.][]{bujarrabal2016, bujarrabal2017, bujarrabal2018, gallardocava2021, gallardocava2023}.
When possible, we used the \trece \unocero\ line, which is the best for our purposes, because this transition is expected to be optically thinner than the \dosuno\ and those of \docep. When only \mbox{\doce\dosuno} was detected, the nebular mass was estimated based on this molecular line. Considering that this line is optically thicker than \trece\unocero, the calculated mass values are underestimated, but can be corrected, as is discussed in \sect\ref{sec:derivedmasses}.
The assumption that the \doce line is optically thicker than that of \trece is valid because, for all sources without $^{13}$CO detection, the $I^{\text{peak}}_{{^{12}}\text{CO}}$/3$\sigma_{^{13}\text{CO}}$ ratio is always much lower than the assumed CO isotopic ratio of 10 (i.e.,\ $^{12}$CO/$^{13}$CO).

For the sources where both lines are detected, we find a \mbox{\doce/\trece} integrated line ratio with a mean value of about 2 for the $J=1-0$ transition, and 6.6 for the ratio between the \doce\dosuno\ and the \trece\unocero\ lines. These values are consistent with those found in other binary post-AGB disks \citep[][]{gallardocava2022}, but they are lower than the ratios typically measured in other post-AGB stars. This suggests that our objects may present a higher \trece abundance, with the binary nature of the sources likely being the origin of such unusually low \mbox{\doce/\trece} ratios.

The derived H$_2$, atomic hydrogen, and helium masses correspond exclusively to the gas regions where CO emission is detected, as CO serves as our tracer for the total gas content.
Contributions from other molecules are assumed to be negligible, since SiO, HCN, SiS, and CS have very low abundances compared to molecular hydrogen, on the order of \xd{-8}\,$-$\,\xd{-9}, \xd{-9}\,$-$\,\xd{-10}, \xd{-11}, and \xd{-10}, respectively \citep[see][for complete description]{gallardocava2022}.

\begin{table*}[t]
	\caption{Nebular mass values derived from the observed $^{12}$CO and $^{13}$CO lines.}
	\vspace{-5mm}
	\begin{center}
		
		\begin{tabular}{lcccccc}
			\hline \hline
			\\[-2ex]
			
			Source   &  $M_{^{12}\text{CO}\, J=2-1}$    & $M_{^{12}\text{CO}\, J=1-0}$ & $M_{^{13}\text{CO}\, J=2-1}$  & $M_{^{13}\text{CO}\, J=1-0}$ & $M_{^{13}\text{CO}\, J=1-0}^{\text{EL}}$ & $d$   \\ 
			&   [M$_{\odot}$]  &   [M$_{\odot}$]   &  [M$_{\odot}$]  &   [M$_{\odot}$] &   [M$_{\odot}$]& [pc]          \\ \hline
			\\[-2ex]
			
			HD\,213985          & 1.8\x\xd{-5}  &  $<$\,6.9\x\xd{-5} & $-$ & $<$\,4.2\x\xd{-4}  & 2.5\x\xd{-4} &  702     \\
			
			RV\,Tau             & 1.8\x\xd{-5}  &  $<$\,1.7\x\xd{-4}  & $-$ & $<$\,1.1\x\xd{-3} & 2.5\x\xd{-4} & 1245   \\ 
			
			HD\,52961           & 8.4\x\xd{-4} & 5.4\x\xd{-3} & 3.7\x\xd{-3} & 2.3\x\xd{-2} &  & 1813    \\ 
			
			U\,Mon              & 1.3\x\xd{-5} & $<$\,8.9\x\xd{-5} &  $<$\,1.9\x\xd{-4} & $<$\,4.7\x\xd{-4} & 1.8\x\xd{-4} & 800   \\
			
			HR\,4049            & 2.6\x\xd{-4} & 6.9\x\xd{-4} & $-$ & 6.7\x\xd{-3}  &     & 1449     \\ 
			
			IRAS\,11472$-$0800  & 1.7\x\xd{-3} & $<$\,3.8\x\xd{-3} & $-$ & $<$\,2.2\x\xd{-2} & 2.3\x\xd{-2} & 6107    \\
			
			\hline
			\\[-2ex]
			
			TW\,Cam             & $<$\,3.6\x\xd{-5} & $<$\,7.2\x\xd{-4} & $-$ & $<$\,5.7\x\xd{-3} & $<$\,4.9\x\xd{-4} & 1757   \\
			IRAS\,05208$-$2035  & $<$\,2.5\x\xd{-5} & $<$\,7.2\x\xd{-4} & $-$ & $<$\,3.5\x\xd{-3} & $<$\,3.4\x\xd{-4} & 1403   \\
			V\,Vul              & $<$\,1.6\x\xd{-5} & $<$\,2.8\x\xd{-4} & $-$ & $<$\,1.8\x\xd{-3} & $<$\,2.2\x\xd{-4} & 1083   \\
			AU\,Peg             & $<$\,1.0\x\xd{-5} & $<$\,1.3\x\xd{-4} & $-$ & $<$\,8.0\x\xd{-4} & $<$\,1.4\x\xd{-4} &  597   \\
			
			\hline

		\end{tabular}
		
	\end{center}
	\small
	\vspace{-1mm}
	\textbf{Notes.}  These mass values were calculated based on the distances indicated in \tab\ref{tab:fuentes_co}. Sources were separated into detections (\textit{top}) and non-detections (\textit{bottom}).
	$M_{^{12}\text{CO}\, J=2-1}$, $M_{^{12}\text{CO}\, J=1-0}$, $M_{^{13}\text{CO}\, J=2-1}$, and $M_{^{13}\text{CO}\, J=1-0}$ indicate the total nebular mass according to the integrated area of the indicated rotational line (see \tab\ref{tab:lineas_detectadas}). $M_{^{13}\text{CO}\, J=1-0}^{\text{EL}}$ is the theorized total nebular mass according to the empirical law that relates the masses derived from the \mbox{$^{13}$CO $J=1-0$} and \mbox{$^{12}$CO $J=2-1$} lines. 
	\label{tab:masas_co}
\end{table*}

\subsection{The derived nebular masses}
\label{sec:derivedmasses}

Based on our CO data (see \tab\ref{tab:lineas_detectadas}), we can estimate the total mass of the pPNe around our binary post-AGB stars using \eq\ref{eq:area_masa}.
We have compiled in \tab\ref{tab:masas_co} the values of the nebular masses derived using each of the rotational lines observed in this work: $M_{^{12}\text{CO}\, J=2-1}$, $M_{^{12}\text{CO}\, J=1-0}$, $M_{^{13}\text{CO}\, J=2-1}$, and $M_{^{13}\text{CO}\, J=1-0}$, according to \mbox{$^{12}$CO $J=2-1$}, \mbox{$^{12}$CO $J=1-0$}, \mbox{$^{13}$CO $J=2-1$}, and \mbox{$^{13}$CO $J=1-0$}, respectively. 
These calculations have been made for both disk sources with molecular CO detection and those without, for which we present upper limits.
 
We begin by presenting the results obtained for the sources HD\,52961 and HR\,4049. These were derived directly from the \trece\unocero\ transition, ensuring that the derived masses are more accurate and representative of the total content in these circumbinary disk-containting post-AGB nebulae.

HD\,52961 presents a nebular mass of 2.3\x\xd{-2}\msp. This source show line profiles with a narrow component and relatively wide wings (see \fig\ref{fig:hd52961_co}). These wings are relatively intense in \doce \dosuno\ and cover a velocity-range of 20\kmsp. We would need high-resolution maps and accurate models to determine the nebular structure and the contribution of the disk to the nebula, allowing us to classify it as either disk-dominated or outflow-dominated.

HR\,4049 presents a total nebular mass of 6.7\x\xd{-3}\msp. This source also exhibits the typical narrow line profiles of rotating gas, further displaying the characteristic double peak of Keplerian disks in its \doce\dosuno\ line profile (see \fig\ref{fig:hr4049_co}).

In the case of HD\,213985, RV\,Tau, U\,Mon, and IRAS\,11472$-$0800, we have only detected the \doce\dosuno\ line. This transition tends to be optically thick (specially in disk-containing nebulae around binary post-AGB stars), so the derived nebular mass would be underestimated. 
Overall, those sources of this work and the previous one \citep[][]{bujarrabal2013a} with well-detected \doce and \trece lines adhere to the following empirical law:
\begin{equation}
\beth = \frac{M_{^{13}\text{CO}\, J=1-0}}{M_{^{12}\text{CO}\, J=2-1}} \approx 13.6.
\end{equation}
Thus, we can provide a more accurate nebular mass values for these four sources based on their assumed \trece \unocero\ emission, the optically thinnest line \citep[in a similar way of][]{bujarrabal2013a}.
In this way, the masses derived from the assumed \trece\unocero\ emission, $M_{^{13}\text{CO}\, J=1-0}^{\text{EL}}$, must be $\beth$ times those derived from the \mbox{$^{12}$CO $J=2-1$} line:
\begin{equation} \label{eq:factor_masa}
M_{^{13}\text{CO}\, J=1-0}^{\text{EL}} = \beth \times  M_{^{12}\text{CO}\, J=2-1} .
\end{equation}
As a result, we present the nebular masses of these objects: 
HD\,213985 has a mass of 2.5\x\xd{-4}\msp, RV\,Tau has 2.5\x\xd{-4}\msp, U\,Mon has 1.8\x\xd{-4}\msp, and IRAS\,11472$-$0800 has 2.3\x\xd{-2}\msp.
These results are good estimates, since \mbox{$M_{^{13}\text{CO}\, J=1-0}^{\text{EL}} \approx M_{^{13}\text{CO}\, J=1-0}$} (see \tab\ref{tab:masas_co}).
These four sources show narrow \doce \dosuno\ line profiles of just $\sim$\,10\kms in range that strongly suggest that the emission from their rotating disks dominate their respective nebulae (see \figs  \ref{fig:213985_co}, \ref{fig:rvtau_co}, \ref{fig:umon_co}, and \ref{fig:i11472_co}). 
High-resolution interferometric observations are necessary to resolve the disk and disk structure. Using the single-dish data and analyses, we estimate only approximate disk masses.

For the non-CO-detected sources, we can estimate upper limits to the nebular mass by using the respective upper limits on the integrated line intensities (see \tab\ref{tab:lineas_detectadas}). 
In a very similar way to those sources detected only in \doce\dosuno, we can estimate their mass upper limit using this same transition, while applying the $\beth$ correction factor for these objects. This approach yields a more stringent constraint on their nebular mass compared to directly using the upper limit derived from the \trece\unocero\ line.
We think that rescaling the \mbox{$^{12}$CO $J=2-1$} result is a more restrictive approach and likely provides a more stringent estimate of the actual mass.
In this way, we find that TW\,Cam has an upper limit to its nebular mass of 4.9\x\xd{-4}\msp, IRAS\,05208$-$2035 of 3.4\x\xd{-4}\msp, V\,Vul of 2.2\x\xd{-4}\msp, and AU\,Peg of 1.4\x\xd{-4}\msp.
These nebular mass upper limits are given in \tab\ref{tab:masas_co}.
The root mean square in the spectra is sufficiently low thanks to the depth of these observations. Despite the non-detection of CO, the nebular mass limit values are considered reliable and consistent with results derived from positive detections.

According to what has been discussed, the most accurate nebular mass values for these CO-emitting circumbinary-disk-containing sources can be found in \tab\ref{tab:todas_masas}. In addition, leveraging the high sensitivity of the observations, upper mass limits for disk objects without molecular emission are listed, and these values enhance the derived nebular mass results.

\section{Diversity and trends in nebular masses}
\label{sec:comparativa}

In this section, we present a comprehensive analysis of all available nebular mass determinations, including both our new measurements and those from previous studies. Our mass determinations, based on both single-dish and interferometric data \citep[][among others]{bujarrabal2013a,bujarrabal2016,bujarrabal2017,bujarrabal2018,gallardocava2021,gallardocava2022mdpi,gallardocava2023}, incorporate two key improvements: (1) updated distance measurements from \citet{bailerjones2021}, and (2) new mass estimates for sources in \citet{bujarrabal2013a} that previously lacked CO detections and upper limits, calculated using the same methodology we apply to non-detections in this study (\sect\ref{sec:metodologia}). Complete details of all mass determinations from previous works are provided in \app\ref{sec:apendice_masas}.

\begin{figure*}[p]
\center
  \centering
	\includegraphics[width=0.90\linewidth]{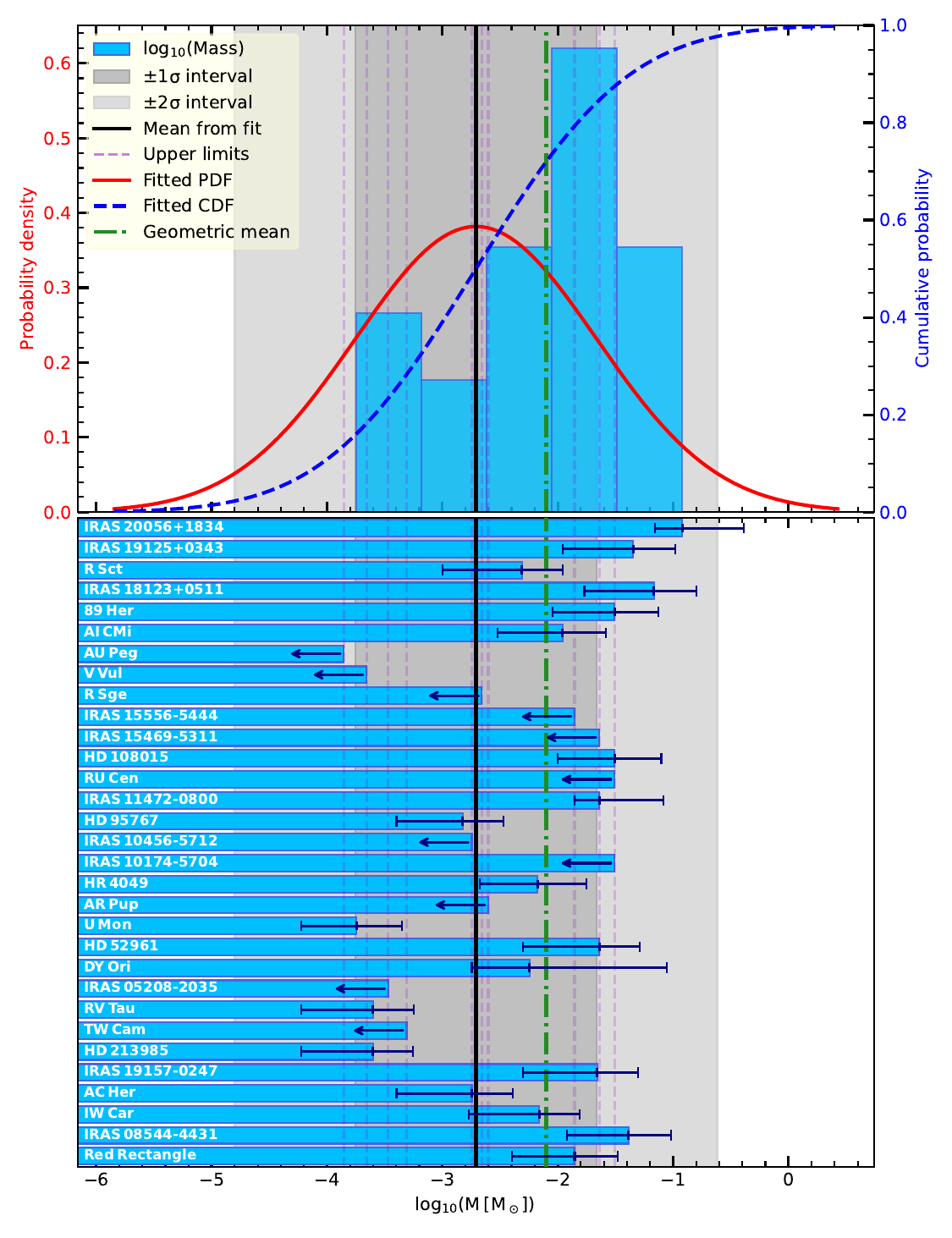}
\caption{\small \textit{Top panel}: Histogram of the logarithm of disk masses for detected sources, overlaid with the probability density function (PDF; red line) and cumulative distribution function (CDF; dashed blue line) resulting from a censored normal likelihood fit that accounts for non-detections (upper limits). Shaded gray regions mark the $\pm$\,1$\sigma_{\text{log}}$ and $\pm$\,2$\sigma_{\text{log}}$ intervals around the fitted mean (vertical black line). The vertical dash-dotted line shows the geometric mean of the detections (green), while vertical dashed lines indicate the positions of the upper limits (in violet).
\textit{Bottom panel}: Representation of individual sources, showing the log-mass of each object. Names appear to the left of each bar. Sources with upper limits are marked with arrows pointing left from their log-mass value. The shaded regions, vertical lines, and color coding are consistent with the top panel. 
}
    \label{fig:histo_masas}  
\end{figure*}

\subsection{Direct measures}
\label{sec:directmeasures}

The full sample of 31 sources observed at millimeter wavelengths to estimate their nebular masses is shown in \fig\ref{fig:histo_masas}. These histograms illustrate the wide variety of results, including both secure detections and upper limits derived from sources without observed molecular emission. 

It is remarkable that this group of disk-containing post-AGB binaries, seemingly similar in nature, displays such a pronounced spread in nebular mass, or in some cases a complete lack of detectable molecular content. This contrast is evident even within subsamples: both disk- and outflow-dominated sources show mass differences exceeding a factor of 20. When the entire sample is considered, including objects with unknown disk-to-outflow mass ratios, the disparity becomes even more dramatic, surpassing a factor of 600.

Beyond this broad dynamic range, the mass distribution of CO-emitting sources is strongly right-skewed. The mean nebular mass is \mbox{$\bar{\mu}_{\text{det}} = 2.3$\x\xd{-2}\msp}, which is significantly higher than the median value of \mbox{$\tilde{\mu}_{\text{det}} = 1.3$\x\xd{-2}\msp}, while the mode, \mbox{$\mu^{*}_{\text{det}} = 8.7$\x\xd{-3}\msp}, closely reflects typical low-mass objects. This pronounced asymmetry, together with a large standard deviation of \mbox{$\sigma_{\text{det}} = 2.9$\x\xd{-2}\msp}, reveals a heavy tail toward higher masses driven by a few extreme outliers (see \tab\ref{tab:mass_statistics}).

In addition, we derived the geometric mean mass for the CO-emitting sources, obtaining \mbox{$\mu_{\text{g, det}} = 8.0$\x\xd{-3}\msp}. This metric offers a more robust characterization of the sample by reducing the impact of extreme values. While the arithmetic mean is strongly affected by high-mass outliers, the geometric mean provides a more representative central tendency for the bulk of the nebular mass distribution, confirming that most systems cluster at lower masses than is indicated by the arithmetic mean. Based on the disk-mass percentage classification by \cite{gallardocava2023}, we find that disk-dominated sources present a geometric mean mass of 1.1\x\xd{-2}\msp, while outflow-dominated sources exhibit a geometric mean mass of 2.9\x\xd{-2}\msp, respectively.
Taken together, these direct statistics highlight the strong diversity in nebular mass among post-AGB binaries, pointing to a wide range of physical conditions, or possibly rapid evolutionary tracks, within this class of systems.

\begin{table}[t]
 \caption{Derived nebular masses for the CO-observed sample of post-AGB disk sources.}
\vspace{-5mm}
\small
\begin{center}

\begin{tabular}{lclc}
\hline \hline
 \\[-2ex]

\multirow{2}{*}{Source} & $M_{\text{neb}}$ & \multirow{2}{*}{Method}  & \multirow{2}{*}{Reference} \\
                & [M$_{\odot}$] &       &            \\
\hline
 \\[-2ex]
Red\,Rectangle       & $ 1.4^{+0.5}_{-0.4} \times 10^{-2}$ & Maps  &  [1] \\
 \\[-2ex]
IRAS\,08544$-$4431   & $ 4.1^{+1.4}_{-1.2}  \times 10^{-2}$ & Maps   & [2] \\
 \\[-2ex]
IW\,Car              & $ 6.9^{+1.7}_{-1.7}  \times 10^{-3}$ & Maps   & [3] \\
 \\[-2ex]
AC\,Her              & $ 1.8^{+0.5}_{-0.4}  \times 10^{-3}$ & Maps   & [4] \\
 \\[-2ex]
IRAS\,19157$-$0247   & $ 2.2^{+0.6}_{-0.5} \times 10^{-2}$ & Single-dish   & [5] \\
 \\[-2ex]
HD\,213985           & $ 2.5^{+0.6}_{-0.6} \times 10^{-4}$ & Single-dish   & [6]  \\
 \\[-2ex]
TW\,Cam              & $\leq\,4.9 \times 10^{-4}$ & Single-dish   & [6] \\
 \\[-2ex]
RV\,Tau              & $ 2.5^{+0.7}_{-0.6} \times 10^{-4}$ & Single-dish  & [6] \\
 \\[-2ex]
IRAS\,05208$-$2035   & $\leq\,3.4 \times 10^{-4}$ & Single-dish   & [6] \\
 \\[-2ex]
DY\,Ori              & $ 5.7^{+76.3}_{-1.8} \times 10^{-3}$ & Single-dish   & [5] \\
 \\[-2ex]
HD\,52961            & $ 2.3^{+0.5}_{-0.5} \times 10^{-2}$ & Single-dish   & [6] \\
 \\[-2ex]
U\,Mon               & $ 1.8^{+0.9}_{-0.6} \times 10^{-4}$ & Single-dish   & [6] \\
 \\[-2ex]
AR\,Pup              & $\leq\,2.5 \times 10^{-3}$ & Single-dish  & [5, 6] \\
 \\[-2ex]
HR\,4049             & $ 6.7^{+4.2}_{-2.1} \times 10^{-3}$ & Single-dish   & [6] \\
 \\[-2ex]
IRAS\,10174$-$5704   & $\leq\,3.1 \times 10^{-2}$ & Single-dish   & [5, 6] \\
 \\[-2ex]
IRAS\,10456$-$5712   & $\leq\,1.8 \times 10^{-3}$ & Single-dish    & [5, 6] \\
 \\[-2ex]
HD\,95767            & $ 1.5^{+0.4}_{-0.4} \times 10^{-3}$ & Single-dish   & [5] \\
 \\[-2ex]
IRAS\,11472$-$0800   & $ 2.3^{+3.6}_{-1.4} \times 10^{-2}$ & Single-dish    & [6] \\
 \\[-2ex]
RU\,Cen              & $\leq\,3.1 \times 10^{-2}$ & Single-dish   & [5, 6] \\
 \\[-2ex]
HD\,108015           & $ 3.1^{+1.7}_{-1.0} \times 10^{-2}$ & Single-dish   & [5] \\
 \\[-2ex]
IRAS\,15469$-$5311   & $\leq\,2.3 \times 10^{-2}$ & Single-dish   & [5, 6] \\
 \\[-2ex]
IRAS\,15556$-$5444   & $\leq\,1.4 \times 10^{-2}$ & Single-dish   & [5, 6] \\
 \\[-2ex]
R\,Sge              & $\leq\,2.2 \times 10^{-3}$ & Single-dish   & [5] \\
 \\[-2ex]
V\,Vul              & $\leq\,2.2 \times 10^{-4}$ & Single-dish   & [6] \\
 \\[-2ex]
AU\,Peg             & $\leq\,1.4 \times 10^{-4}$ & Single-dish    & [6] \\
 \\[-2ex]
AI\,CMi             & $ 1.1^{+0.4}_{-0.3} \times 10^{-2}$ & Single-dish    & [5] \\
 \\[-2ex]
89\,Her             & $ 3.1^{+1.2}_{-0.9} \times 10^{-2}$ & Maps    & [7] \\
 \\[-2ex]
IRAS\,18123+0511    & $ 6.8^{+2.4}_{-1.7} \times 10^{-2}$ & Single-dish    & [5] \\
 \\[-2ex]
R\,Sct              & $ 4.9^{+1.3}_{-1.0} \times 10^{-3}$ & Maps   & [4] \\
 \\[-2ex]
IRAS\,19125+0343    & $ 4.5^{+1.5}_{-1.0} \times 10^{-2}$ & Maps   & [4] \\
 \\[-2ex]
IRAS\,20056+1834    & $ 1.2^{+1.7}_{-0.7} \times 10^{-1}$ & Single-dish   & [5] \\
 \\[-2ex]
\hline
\end{tabular}

\end{center}
\small
\vspace{-1mm}
\textbf{Notes.} 
This study provides the most up-to-date compilation for all CO-observed sources, unifying the results presented here with prior findings and employing the distance estimates from \citet{bailerjones2021}.
References: [1] \citet{bujarrabal2016}; [2] \citet{bujarrabal2018}; [3] \citet{bujarrabal2017}; [4] \citet{gallardocava2021}; [5] \citet{bujarrabal2013a}; [6] this work; [7] \citet{gallardocava2023}. For sources that have two different references, see \app\ref{sec:velocidades}.

\label{tab:todas_masas}
\end{table}

\subsection{Normal distribution of nebular mass}

To extend our mass distribution analysis, we incorporated upper limits by assuming a normal distribution for the logarithm of the masses (log-norm from now on), which is empirically justified by the relatively symmetric spread of log$_{10}$-transformed detected masses (see \fig\ref{fig:histo_masas}\,\textit{top}) and the consistent trend of upper limits with the observed distribution.

A log-normal distribution is well suited for this dataset, as the masses span multiple orders of magnitude and are strictly positive, effectively capturing the multiplicative nature of mass formation and the skewness of the data \citep[][]{limpert2001}. Our maximum likelihood approach incorporates the full dataset: for detections we include asymmetric mass uncertainties through distribution convolution, while for non-detections we properly account for upper limits using censored data analysis. This comprehensive treatment prevents bias and provides an accurate estimate of the underlying mass distribution for the entire population of CO-detected and non-CO-detected sources. In \fig\ref{fig:histo_masas}\,\textit{top}, we present the resulting log-normal fit.

This approach yields a log-normal distribution with parameters $\mu_{\text{log}} = -2.7$ and $\sigma_{\text{log}} = 1.0$ (see \tab\ref{tab:mass_statistics}).
This fit reveals fundamental characteristics of the population that are not apparent from direct detections alone. The corresponding median mass, $\tilde{\mu}_{\text{log}} = 2.0 \times 10^{-3}$\msp, agrees well with the median of detected values, suggesting that it robustly represents the typical object in the overall disk-containing population. However, the log-normal model provides richer information on the shape of the distribution. The $\pm$\,1$\sigma_{\text{log}}$ interval, $[1.8 \times 10^{-4}, 2.2 \times 10^{-2}]$\msp, shows that 68\% of sources fall within nearly an order of magnitude, while the broader $\pm$\,2$\sigma_{\text{log}}$ range, $[1.6 \times 10^{-5}, 2.4 \times 10^{-1}]$\msp, highlights the asymmetry of the distribution, with a long tail toward higher masses. This explains why the arithmetic mean of detected sources, $\bar{\mu}_{\text{det}} = 2.3 \times 10^{-2}$\msp, is skewed upward relative to the median, as it is sensitive to massive outliers (see \tab\ref{tab:mass_statistics}).

When comparing these results with the statistics from direct detections only, several important differences emerge. The detection-only arithmetic mean is notably higher than the median of the log-normal fit, reflecting how the inclusion of censored data affects population estimates. The geometric mean of detections ($\mu_{\text{g, det}} = 8.0 \times 10^{-3}$\msp), being less sensitive to extreme values, lies between the log-normal median and the detection mean, further highlighting the positive skewness of the distribution. The standard deviation of direct detections (\mbox{$\sigma_{\text{det}} = 2.9 \times 10^{-2}$\msp}), while large, underestimates the full spread implied by the log-normal standard deviation apparent in the sigma parameter of the log-normal fit and its associated confidence intervals. Most importantly, the detection-only analysis misses the true dynamic range of the population, particularly the substantial number of low-mass objects implied by the lower bounds of the log-normal fit.

\begin{table}[t]
 \caption{Statistical properties of the nebular mass distribution.}
\vspace{-5mm}
\begin{center}

\begin{tabular}{lc}
\hline \hline \\[-2ex]
\multicolumn{2}{c}{Statistics on detections only} \\
\hline \\[-2ex]
Arithmetic mean, $\bar{\mu}_{\text{det}}$\,[M$_{\odot}$]               & $2.3 \times 10^{-2}$  \\
Geometric mean, $\mu_{\text{g, det}}$\,[M$_{\odot}$]     & $8.0 \times 10^{-3}$  \\
Median, $\tilde{\mu}_{\text{det}}$\,[M$_{\odot}$]                  & $1.3 \times 10^{-2}$  \\
Mode, $\mu^{*}_{\text{det}}$\,[M$_{\odot}$]                      & $8.7 \times 10^{-3}$  \\
Standard deviation, $\sigma_{\text{det}}$\,[M$_{\odot}$]          & $2.9 \times 10^{-2}$  \\
\hline
 \\[-2ex]
\multicolumn{2}{c}{Base-10 logarithm fit with censoring} \\
\hline \\[-2ex]
Mean, $\bar{\mu}_{\text{log}}$                  & $-2.7$ \\
Standard deviation, $\sigma_{\text{log}}$      & $1.0$ \\
Median, $\tilde{\mu}_{\text{log}}$\,[M$_{\odot}$]                         & $2.0 \times 10^{-3}$ \\
$\pm$\,$1\sigma_{\text{log}}$ interval \,[M$_{\odot}$]     & $[1.8 \times 10^{-4},\,2.2 \times 10^{-2}]$ \\
$\pm$\,$2\sigma_{\text{log}}$ interval \,[M$_{\odot}$]    & $[1.6 \times 10^{-5},\,2.4 \times 10^{-1}]$ \\
\hline
\end{tabular}

\end{center}
\small
\vspace{-1mm}
\textbf{Notes.} 
Statistical summary of the nebular mass distribution. The first section shows statistics computed only for detections, including the arithmetic mean, geometric mean, median, mode, and standard deviation in linear scale (masses in solar units). The second section presents the parameters from a censored log-normal fit, where $\bar{\mu}_{\text{log}}$ and $\sigma_{\text{log}}$ are the mean and standard deviation of the base-10 logarithm of the mass, respectively. Also included are the median $\tilde{\mu}_{\text{log}}$ and the $\pm$\,$\sigma_{\text{log}}$ and $\pm$\,2$\sigma_{\text{log}}$ confidence intervals in solar mass units.
See \fig\ref{fig:histo_masas}.
\label{tab:mass_statistics}
\end{table}

These differences have important implications for interpreting the post-AGB disk population. The log-normal fit reveals that most objects have relatively low masses clustered around the median, along with a smaller subpopulation of significantly more massive disks that skew the arithmetic mean upward. This explains why the mean derived from detections alone exceeds both the median and the geometric mean, reflecting a bias toward more easily detectable, massive sources in the observational sample.
By integrating both detections and upper limits into a statistically consistent framework, the log-normal model offers a more complete and accurate characterization of the underlying mass distribution. It shows that the majority of objects are less massive than direct detections suggest, and that relying solely on simple summary statistics underestimates both the spread and central tendency of the data.
The consistency in median values across methods supports the robustness of this parameter, while the divergence in other metrics underscores the need for rigorous statistical treatment in studies of circumbinary disk-bearing pPNe. Ultimately, the results provide compelling evidence that the mass distribution in these systems is highly asymmetric, shaped by a substantial population of low-mass, often undetected, disks.

\subsection{Uncertainties}

We are well aware that there is a nebular mass problem.
What we derive can be considered to be a lower limit to the total mass lost by the primary, as the derived values are very low when compared to the amount of mass the star must have lost to become a post-AGB star (however, this may not be the case for post-RGB primaries). This missing mass could reside in non-molecular components. In \citet{gallardocava2023}, the possible contributions from atomic material, the collimated low-density high-velocity jets launched by the companion during the late AGB and early post-AGB stages, and the frequently debated presence of very extended halos originating during the AGB phase, were analyzed. It was concluded that these non-molecular
structures are unlikely to significantly contribute to the total nebular mass.

In our targets, one possibility can be that a significant fraction was accreted by the companion star during or after the mass-loss phase of the primary, potentially accelerating its evolution. This mass-loss problem is not exclusive to binary post-AGB stars. It is also present in many evolved nebulae \citep[][]{santandergarcia2022}, where we are not able to determine where this missing nebular mass resides.

The uncertainties associated with the calculated masses were determined by considering that the main sources of error are the absolute calibration uncertainty of the observations and the distance to each source.
For the absolute calibration error, we found values of approximately 10\% and 15\% for the $J=1-0$ and $J=2-1$ transitions, respectively.
Regarding distances, this parameter significantly contributes to the overall uncertainties due to the inverse-square dependence of the mass on distance. The asymmetry in their uncertainties propagates into similarly asymmetric errors in our results (see \tab\ref{tab:todas_masas}).

\section{Searching for molecules other than CO}
\label{sec:othermolecules}

\subsection{Observational results}

The CO-emitting sources analyzed in this work have been extensively observed in the \textit{Q} band using the Yebes-40\,m telescope, as part of a dedicated effort to investigate their molecular content beyond the standard CO lines. Complementary observations were also carried out with the IRAM-30\,m telescope during the CO campaigns.

The primary objective of the Yebes-40\,m \textit{Q}-band observations was the search for SiO maser emission, along with the detection of the SiO thermal line. Despite the depth and sensitivity of our observational setup, no SiO emission, either maser or thermal, was detected in any of the targeted sources. This result, however, is consistent with expectations: previous comprehensive surveys already indicated the molecular scarcity that characterizes these post-AGB nebulae hosting rotating disks \citep[see][for details]{gallardocava2022}.

Even in the absence of clear detections, the non-detections themselves are scientifically valuable. The stringent upper limits derived from these observations provide meaningful constraints on the presence of SiO. 
The derived upper limits for each source are summarized in \tab\ref{tab:qband}.

\subsection{Lack of molecular content}

Here, we discuss several possible reasons that might explain the low or apparent absence of molecular content in some of these circumbinary-disk-containing objects.

One possible reason for the apparent lack of CO content in these sources could be the presence of very low-mass rotating disks. In this case, future, deeper, and more sensitive observations might detect weak molecular emission, revealing these faint rotating disks. Additionally, the absence of CO emission in these sources could be attributed to photodissociation effects. 
Although our post-AGB stars are not hot enough efficiently photodissociate CO, the binary system can still destroy molecules in the inner regions of the nebula due to the UV radiation from the accretion disk around the companion or the interaction with the ejected ionized wind. Additionally, CO can be subject to freeze-out and interstellar-UV photodissociation in the outer and colder parts of the rotating disk. Another possibility for explaining this poor molecular content is that the rotating disk is not detected simply because it is too small. 
However, this scenario is unlikely, since for the disk to be both massive and undetected it would need to be optically thick, which would actually make it more detectable due to the higher densities.

The absence of molecular content can also be observed in protoplanetary disks. The aforementioned phenomena explaining the lack of CO have been widely confirmed in these disks around young stars
\citep[see][and references therein]{pietu2014,barenfeld2017,miotello2021}.

Another more plausible possibility is that CO escapes from the disk through outflows, while the dust, particularly large grains, remains gravitationally bound (Gallardo Cava et al. in prep). In this scenario, the CO non-detections may correspond to older disks that have experienced prolonged molecular gas depletion due to sustained outflow activity.

In summary, it would be possible to differentiate between very low-mass rotating disks and very high-density compact disks. For both cases, and even for those showing CO emission, the additional effect of photodissociation occurring in both the inner and outer regions of the disk must be considered.


\begin{figure}[t]
\center
\includegraphics[width=\sz\linewidth]{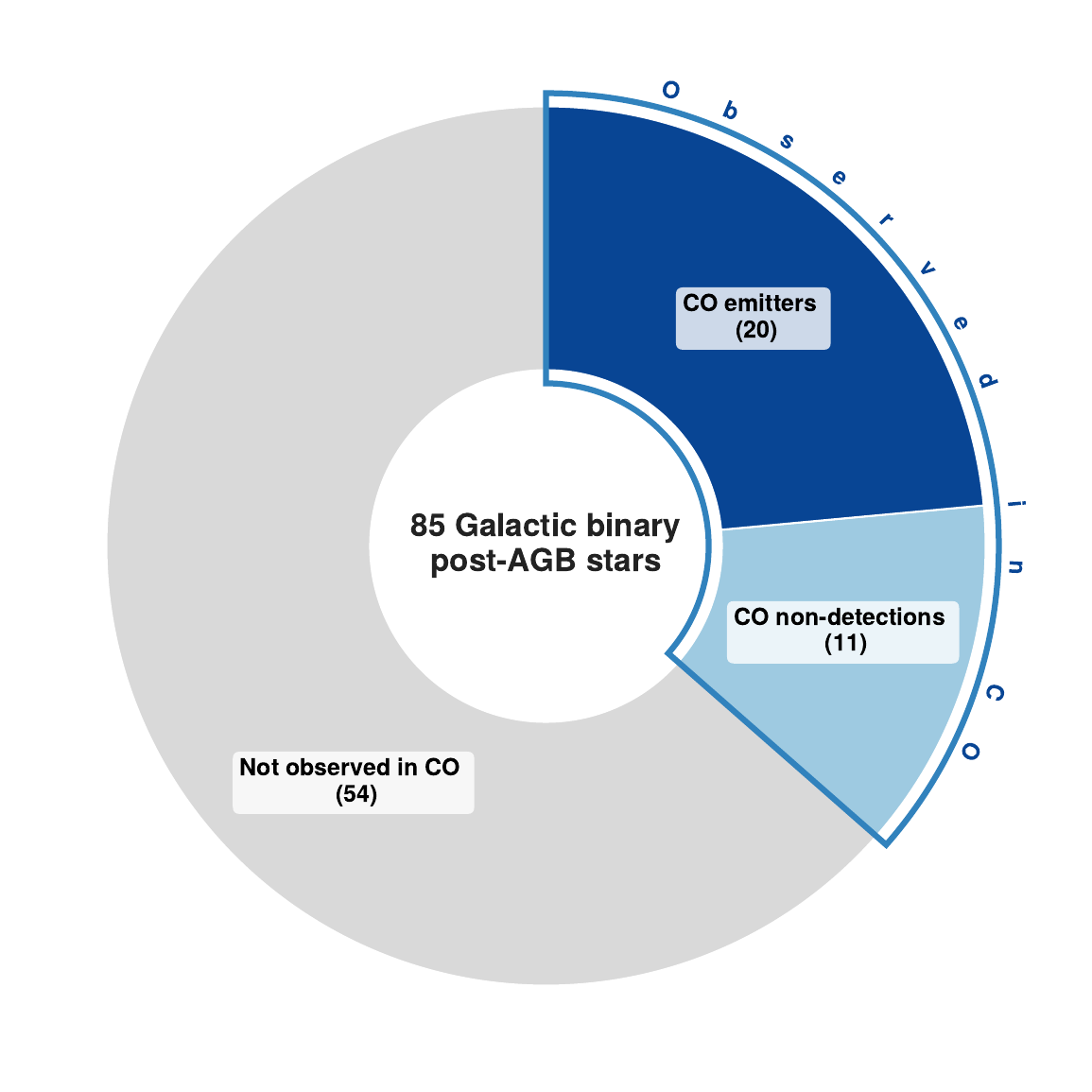} 
\caption{\small 
Distribution of 85 known Galactic binary post-AGB stars based on CO observations. The gray segment represents stars not observed in CO. Among those observed (encircled in blue), the dark blue sector corresponds to CO emitters, while the light blue sector shows non-detections.
}
    \label{fig:quesitos}  
\end{figure}

\section{Conclusions}
\label{sec:conclusiones}

To date, 85 Galactic post-AGB binary objects with circumbinary disks (studied through optical and infrared observations) are known \citep[][]{kluska2019}.
In this work, we have observed ten sources in mm-waves through single-dish observations using the IRAM-30\,m radiotelescope (see \tab\ref{tab:fuentes_co}).
These new sources significantly expand the total sample of objects observed in mm-wave in the search of CO.
We note that three of the objects in this study were previously observed in the earlier survey; however, these new observations are much deeper and sensitive, leading to an increase in the percentage of objects studied at these wavelengths from 28\% to 37\%.

In this work, we have confirmed that six out of the ten studied objects do emit in CO: HD\,213985, RV\,Tau, HD\,52961, U\,Mon, HR\,4049, and IRAS\,11472$-$0800. These six new detections significantly increase the sample of objects that emit in CO, rising from 18\% to 25\% of the 85 disk-containing post-AGB nebulae. 
Regarding the other four new objects that do not emit in molecules in a detectable way, TW\,Cam, IRAS\,05208$-$2035, V\,Vul, and AU\,Peg, despite the absence of CO emission, we have achieved very promising upper limits.
The main parameters derived from the observations presented in this work can be consulted in \tab\ref{tab:lineas_detectadas}.
Of the 31 objects studied in mm-waves \citep[][and this work]{bujarrabal2013a, bujarrabal2016, bujarrabal2017, bujarrabal2018, gallardocava2021, gallardocava2023}, 20 show CO emission, representing a 65\% success rate in molecular detection (see \fig\ref{fig:quesitos}).

From these new CO data, we calculated the derived nebular mass for each of the observed rotational lines (see \tab\ref{tab:masas_co}). We observe that the mass increases as the studied line becomes less opaque. Therefore, we used the \trece\unocero\ data whenever possible. However, we find that for all disk-containing sources emitting in both \doce\ and \trecep, the ratio of derived nebular masses from the \mbox{\trece\unocero} and \doce\dosuno\ lines is 13.6. Thus, for sources with \doce detections but without \trece\ emission, we have provided a more accurate estimate of the total nebular mass corresponding to its presumed emission in this molecule, which is reasonably consistent with the upper limits obtained. For sources without molecular emission, we have calculated upper limits for the nebular mass for each of the observed rotational lines.
Our best-estimated nebular mass values for this sample of ten objects can be found in \tab\ref{tab:todas_masas}.

Moreover, we have compared the nebular mass values from this work with all other existing values from previous studies using the most recent distances. According to \fig\ref{fig:histo_masas}, the data show a normal distribution when examined on a logarithmic scale with high variability, as these types of disk-containing sources exhibit a wide range of total nebular mass values (including both the rotating and expanding components). This dispersion is evident even in sources where the contribution of the disk to the total nebular mass is similar. However, overall, we can find mass differences that exceed a factor of 600.
The full CO-studied dataset presented in \tab\ref{tab:todas_masas} constitutes the most comprehensive and up-to-date compilation of nebular masses for post-AGB sources.

Additionally, we have presented a fit of our results to a log-normal distribution. This modeling approach incorporates the upper limits derived for the entire sample, recognizing their statistical and scientific value.
However, this model exhibits a somewhat larger dispersion, as it inherently gives more weight to extreme values. Despite this limitation, the log-normal approach offers a significant advantage: it accounts for the rotating disks for which we only have upper limits to the mass, ensuring that these disk-containing sources are not excluded from the overall statistical analysis.

The typical mass of these objects, accounting for both detections and non-detections through censored data analysis, is approximately 2.0\x\xd{-3}\msp. This value corresponds to the median of the log-normal fit, making it a robust indicator that is not biased by the presence of high-mass outliers.

The CO-emitting sources analyzed in this work have been deeply observed in the \textit{Q} band with the aim of detecting molecular species beyond CO, with particular focus on SiO maser and thermal emission. Despite the sensitivity of our observations, no SiO emission was detected in any of the targets. This lack of detection reinforces the idea that these disk-bearing post-AGB objects are severely deficient in molecular content, an intriguing characteristic whose origin remains an open question.
A plausible explanation for the CO non-detections and lack of other molecules is the preferential removal of molecular gas through outflows, while the dust, particularly larger grains, remains gravitationally bound, leading to compact disk sources. This scenario implies that some of the undetected systems may be more evolved disks that have experienced prolonged gas depletion over time.

Finally, this work lays a valuable foundation for future research, as it paves the way for detailed investigations into the intrinsic nature of these systems. Although these systems appear remarkably similar at first glance, they exhibit substantial differences across various physical and observational properties. The only aspect we can say with certainty that they share is the presence of an evolved binary central star and a surrounding circumbinary disk.

\begin{acknowledgements}
We are grateful to the anonymous referee for the valuable recommendations and comments.
This work is part of the coordinated research project NEBULAe WEB, grants PID2019-105203GB-C21 funded by MICIU/AEI/10.13039/501100011033, and coordinated project MESON, grant PID2023-146056NB-C21, funded by MICIU/AEI/10.13039/501100011033 and ERDF/EU.
This work is based on observations carried out under project number 144-17 with the 30\,m telescope. IRAM is supported by INSU/CNRS (France), MPG (Germany) and IGN (Spain).
IRAM, the Institut de radioastronomie millimétrique, is an international research institute funded the French Centre National de la Recherche Scientifique (CNRS), the German Max-Planck Gesellschaft (MPG) and the Spanish Instituto Geográfico Nacional (IGN). 
This work is also based on observations carried out with the Yebes-40\,m telescope (20B006). The 40\,m radiotelescope at Yebes Observatory is operated by the Spanish Geographic Institute (IGN; Ministerio de Transportes y Movilidad Sostenible). 
This work has made use of GILDAS software for data reduction and analysis (https://www.iram.fr/IRAMFR/GILDAS).

\end{acknowledgements}

\bibliographystyle{aa}
\bibliography{referencias}

\appendix

\section{Additional nebular mass calculations}
\label{sec:apendice_masas}

\begin{table}[h]
 \caption{Nebular mass values derived from previous single-dish CO observations without detections.}
\vspace{-5mm}
\begin{center}

\begin{tabular}{lccc}
\hline \hline
 \\[-2ex]
\multirow{2}{*}{Source}   & $\int I\,\text{d}V$    & $M_{\text{neb}}$    & $d$   \\ 
                          & [K\,km\,s$^{-1}$]      & [M$_{\odot}$]       & [pc]  \\ \hline
 \\[-2ex]
 
 AR\,Pup                  & $\leq$\,4.8\x\xd{-1}   & $<$\,2.5\x\xd{-3}    & 661    \\
 IRAS\,10174$-$5704       & $\leq$\,1.6\x\xd{-1}   & $<$\,3.1\x\xd{-2}    & 850    \\ 
 IRAS\,10456$-$5712       & $\leq$\,1.2\x\xd{-1}   & $<$\,1.8\x\xd{-3}    & 1104   \\ 
 RU\,Cen                  & $\leq$\,1.2\x\xd{-1}   & $<$\,3.1\x\xd{-2}    & 4618   \\  
 IRAS\,15469$-$5311       & $\leq$\,1.6\x\xd{-1}   & $<$\,2.3\x\xd{-2}    & 3471   \\ 
 IRAS\,15556$-$5444       & $\leq$\,1.2\x\xd{-1}   & $<$\,1.4\x\xd{-2}    & 3105   \\

\hline

\end{tabular}

\end{center}
\small
\vspace{-1mm}
\textbf{Notes.} 
The calculations have been performed based on \doce\dosuno\ data obtained from \citet{bujarrabal2013a} using the distances of \citet{bailerjones2021}.
Integrated emission (area) units are expressed in K\,km\,s$^{-1}$, with Kelvin indicated in main beam temperature units.
\label{tab:masas_co_2013}
\end{table}

In this appendix, we present the derived mass values for those seven sources presented in \fig\ref{fig:histo_masas} that were observed in \citet{bujarrabal2013a}, and for which only the observational $\sigma$ was provided without an estimation of the upper limit for the nebular mass. 
These objects were observed in the \doce\dosuno\ and \doce\tresdos\ lines with the APEX telescope.

Analogous to the method used for the sources in this work, we first calculate an upper limit for the integrated area using \eq\ref{eq:incertidumbre_area}. In this case, the number of channels is again the average of the channels that typically covered the observed line, while the remaining values can be found in the original article. The method for calculating the mass is the same as described in \sect\ref{sec:metodologia}. In this way, the masses derived from the \doce\dosuno\ line should be corrected by multiplying them by a factor of 13.6 to match the masses derived from the assumed \trece\unocero\ emission. 

The upper limits to the nebular mass for each of these sources, along with the upper limit of the integrated area of the \doce\dosuno\ line used to calculate the mass, can be found in \tab\ref{tab:masas_co_2013}.

\section{Heliocentric velocity comparison}
\label{sec:velocidades}

In this appendix, we present a comparison between the heliocentric velocities derived from our CO spectral line observations and the systemic velocities previously reported \citep[][]{oomen2018}.

\begin{figure}[h]
\center
\includegraphics[width=1.08\linewidth]{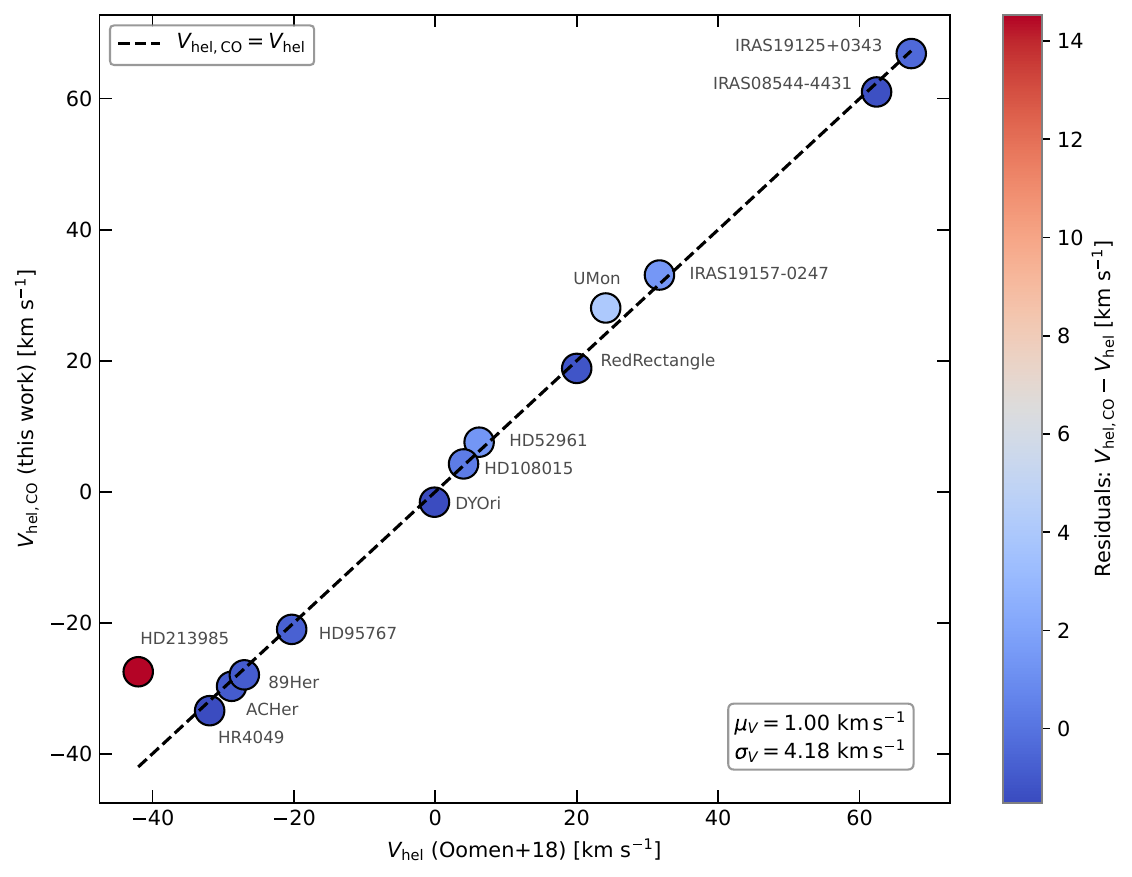} 
\caption{\small 
Comparison between heliocentric velocities derived from CO observations, $V_{\mathrm{hel, CO}}$, and those reported by \citet{oomen2018}, $V_{\text{hel}}$. Each point corresponds to an individual disk-containing source, with the color indicating the residual velocity difference ($V_{\text{hel, CO}} - V_{\text{hel}}$). The dashed line represents the one-to-one relation ($V_{\mathrm{hel, CO}} = V_{\mathrm{hel}}$). The average offset $\mu_{\text{V}}$ and scatter $\sigma_{\text{V}}$ of the residuals are shown in the bottom-right box. 
}
    \label{fig:comparativa_v}  
\end{figure}

In most cases, the velocities derived from our CO observations are in excellent agreement with those reported by \citet{oomen2018}, showing very low dispersion (see \tab\ref{tab:comparativa_v_tabla} and \fig\ref{fig:comparativa_v}). The mean $\mu_{\text{V}}$ and scatter $\sigma_{\text{V}}$ of the residuals further confirm the strong consistency between both datasets. This agreement reinforces the reliability of the velocity measurements in tracing the systemic motion of post-AGB objects.

One notable exception is HD\,213985, for which the CO-derived velocity significantly deviates from the value listed by \citet{oomen2018}. The \doce\dosuno\ line profile observed for this source is very narrow and symmetric, characteristic of emission from a Keplerian rotating disk (see \fig\ref{fig:213985_co}).
We are confident that the emission does not originate from ISM features. In typical ISM conditions, the lower-energy \doce\unocero\ transition is significantly stronger than \doce\dosuno\ \citep[e.g.,][]{falgarone1998}. In our observations, the \doce\unocero\ line is not detected, while the \doce\dosuno\ line is clearly present, which is incompatible with an ISM origin and supports a circumstellar disk scenario.

\begin{table}[t]
 \caption{Comparison between heliocentric velocities derived from CO observations and those compiled by \citet{oomen2018} for the CO-observed post-AGB disk sources.}
\vspace{-5mm}
\begin{center}

\begin{tabular}{lcccc}
\hline \hline 

\multirow{2}{*}{Source} & $V_{\mathrm{LSR}}$ & $V_{\mathrm{hel,\, CO}}$ & $V_{\mathrm{hel}}$ & Residual \\
 & [km\,s$^{-1}$] & [km\,s$^{-1}$] & [km\,s$^{-1}$] & [km\,s$^{-1}$] \\
\hline
 \\[-2ex]

Red\,Rectangle & 0.0 & 18.8 & 20.0 & $-$1.2 \\
IRAS\,08544$-$4431 & 45.0 & 61.0 & 62.4 & $-$1.4 \\
AC\,Her & $-$10.0 & $-$29.69 & $-$28.8 & $-$0.9 \\
IRAS\,19157$-$0247 & 49.0 & 33.1 & 31.7 & 1.4 \\
HD\,213985 & $-$29.0 & $-$27.5 & $-$42.0 & 14.5 \\
DY\,Ori & $-$16.0 & $-$1.6 & $-$0.1 & $-$1.5 \\
HD\,52961 & $-$7.0 & 7.6 & 6.2 & 1.4 \\
U\,Mon & 10.5 & 28.1 & 24.1 & 3.9 \\
HR\,4049 & $-$45.0 & $-$33.4 & $-$31.9 & $-$1.5 \\
HD\,95767 & $-$32.0 & $-$21.0 & $-$20.3 & $-$0.7 \\
HD\,108015 & $-$2.0 & 4.3 & 4.0 & 0.3 \\
89\,Her & -8.0 & $-$27.9 & $-$27.0 & $-$0.9 \\
IRAS\,19125+0343 & 84.0 & 66.9 & 67.3 & $-$0.4 \\

\hline
\end{tabular}

\end{center}
\small
\vspace{-1mm}
\textbf{Notes.} 
The heliocentric velocities derived from CO emission ($V_{\text{hel,\,CO}}$) are compared with those compiled by \citet{oomen2018} ($V_{\text{hel}}$), and the residuals are defined as $V_{\text{hel,\,CO}} - V_{\text{hel}}$. The LSR velocities ($V_{\text{LSR}}$) are included for reference. Velocities are in km\,s$^{-1}$. The CO velocities can be found in this work and in \citet{bujarrabal2013a, bujarrabal2016, bujarrabal2017, bujarrabal2018, gallardocava2021}.
\label{tab:comparativa_v_tabla}
\end{table}

\section{Yebes-40\,m and IRAM-30\,m results}

In this appendix, we present the results for the most relevant lines for those sources of this work in which CO lines have been detected: SiO (both thermal and maser emission), $^{29}$SiO, $^{30}$SiO, CS, SO, C$^{17}$O, and C$^{18}$O. The results are summarized in the following table (see \tab\ref{tab:qband}):

\begin{table*}[t]
\caption{Summary of the most relevant lines observed using both the Yebes-40\,m and  IRAM-30\,m radiotelescopes.}
\tiny
\vspace{-5mm}
\begin{center}

\begin{tabular*}{\textwidth}{@{\extracolsep{\fill\quad}}lllcccccccl}
\hline \hline
\noalign{\smallskip}

\multirow{2}{*}{Source} & \multirow{2}{*}{Molecule} & \multirow{2}{*}{Transition} & $I$\,(peak) & $\sigma$ & $\int I\,\text{d}V$ & $\sigma \left(\int I\,\text{d}V \right)$ & $\delta V$ & $V_{\text{LSR}}$ & \multirow{2}{*}{Comments} \\
& & & [Jy] & [Jy] & [Jy\,km\,s$^{-1}$] & [Jy\,km\,s$^{-1}$] & [km\,s$^{-1}$] & [km\,s$^{-1}$] & \\
\hline
\\[-2ex]

HD\,213985 & SiO        & $v=0 \ J=1-0$    &  $\leq$\,3.5E-02  &   1.2E-02  &  $\leq$\,9.5E-02  &   3.2E-02  &  1.1     &        $-$ & Not detected   \\
           &            & $v=1 \ J=1-0$    &  $\leq$\,3.0E-02  &   1.0E-02  &  $\leq$\,8.2E-02  &   2.7E-02  &  1.1     &        $-$ & Not detected   \\
           &            & $v=2 \ J=1-0$    &  $\leq$\,2.3E-02  &   7.8E-03  &  $\leq$\,6.3E-02  &   2.1E-02  &  1.1     &        $-$ & Not detected   \\
           & $^{29}$SiO & $v=0 \ J=1-0$    &  $\leq$\,2.5E-02  &   8.4E-03  &  $\leq$\,6.8E-02  &   2.3E-02  &  1.1     &        $-$ & Not detected   \\
           & $^{30}$SiO & $v=0 \ J=1-0$    &  $\leq$\,2.1E-02  &   6.9E-03  &  $\leq$\,5.6E-02  &   1.9E-02  &  1.1     &        $-$ & Not detected   \\
           & CS         & $v=0 \ J=1-0$    &  $\leq$\,5.8E-02  &   1.9E-02  &  $\leq$\,1.3E-01  &   4.3E-02  &  0.9     &        $-$ & Not detected   \\

\hline     
\\[-2ex]  

RV\,Tau   & SiO        & $v=0 \ J=1-0$     &   $\leq$\,2.5E-02  &  8.4E-03  &  $\leq$\,9.2E-02  &  3.1E-02  &  1.1  &        $-$ & Not detected   \\
          &            & $v=1 \ J=1-0$     &   $\leq$\,2.7E-02  &  9.0E-03  &  $\leq$\,9.9E-02  &  3.3E-02  &  1.1  &        $-$ & Not detected   \\
          &            & $v=2 \ J=1-0$     &   $\leq$\,2.0E-02  &  6.5E-03  &  $\leq$\,7.2E-02  &  2.4E-02  &  1.1  &        $-$ & Not detected   \\
          & $^{29}$SiO & $v=0 \ J=1-0$     &   $\leq$\,2.3E-02  &  7.6E-03  &  $\leq$\,8.4E-02  &  2.8E-02  &  1.1  &        $-$ & Not detected   \\
          & $^{30}$SiO & $v=0 \ J=1-0$     &   $\leq$\,2.1E-02  &  6.9E-03  &  $\leq$\,7.5E-02  &  2.5E-02  &  1.1  &        $-$ & Not detected   \\
          & CS         & $v=0 \ J=1-0$     &   $\leq$\,4.5E-02  &  1.5E-02  &  $\leq$\,1.3E-01  &  4.5E-02  &  0.9  &        $-$ & Not detected   \\

\hline     
\\[-2ex] 

HD\,52961 & SiO        & $v=0 \ J=1-0$     & $\leq$\,1.8E-02 & 6.0E-03 & $\leq$\,4.5E-02 & 1.5E-02 & 0.5 & $-$ & Not detected   \\
          &            & $v=1 \ J=1-0$     & $\leq$\,1.9E-02 & 6.4E-03 & $\leq$\,4.8E-02 & 1.6E-02 & 0.5 & $-$ & Not detected   \\
          &            & $v=2 \ J=1-0$     & $\leq$\,1.9E-02 & 6.4E-03 & $\leq$\,4.8E-02 & 1.6E-02 & 0.5 & $-$ & Not detected   \\
          & $^{29}$SiO & $v=0 \ J=1-0$     & $\leq$\,1.9E-02 & 6.5E-03 & $\leq$\,4.8E-02 & 1.6E-02 & 0.5 & $-$ & Not detected   \\
          & $^{30}$SiO & $v=0 \ J=1-0$     & $\leq$\,2.1E-02 & 7.0E-03 & $\leq$\,5.3E-02 & 1.8E-02 & 0.5 & $-$ & Not detected   \\
          & CS         & $v=0 \ J=1-0$     & $\leq$\,3.5E-02 & 1.2E-02 & $\leq$\,8.7E-02 & 2.9E-02 & 0.9 & $-$ & Not detected   \\

\hline     
\\[-2ex]

U\,Mon    & SiO        & $v=0 \ J=1-0$     &  $\leq$\,2.9E-02  &  9.5E-03  &  $\leq$\,1.2E-01  & 4.1E-02  &  1.1   & $-$ & Not detected   \\
          &            & $v=0 \ J=5-4$     &  $\leq$\,2.4E-01  &  7.8E-02  &  $\leq$\,4.6E-01  & 1.5E-01  &  0.5   & $-$ & Not detected   \\
          &            & $v=1 \ J=1-0$     &  $\leq$\,2.8E-02  &  9.3E-03  &  $\leq$\,1.2E-01  &  4.0E-02  &  1.1  & $-$ & Not detected   \\
          &            & $v=2 \ J=1-0$     &  $\leq$\,2.6E-02  &  8.8E-03  &  $\leq$\,1.1E-01  &  3.8E-02  &  1.1  & $-$ & Not detected   \\
          & $^{29}$SiO & $v=0 \ J=1-0$     &  $\leq$\,2.4E-02  &  7.9E-03  &  $\leq$\,1.0E-01  &  3.4E-02  &  1.1  & $-$ & Not detected   \\
          & $^{30}$SiO & $v=0 \ J=1-0$     &  $\leq$\,2.2E-02  &  7.4E-03  &  $\leq$\,9.5E-02  &  3.2E-02  &  1.1  & $-$ & Not detected   \\
          & CS         & $v=0 \ J=1-0$     &  $\leq$\,5.5E-02  &  1.8E-02  &  $\leq$\,1.9E-01  &  6.4E-02  &  0.9  & $-$ & Not detected   \\
          & C$^{17}$O  & $v=0 \ J=2-1$     &  $\leq$\,8.0E-02  &  2.4E-01  &  $\leq$\,4.7E-01  &  1.6E-01  &  0.5  & $-$ & Not detected   \\
          & C$^{18}$O  & $v=0 \ J=2-1$     &  $\leq$\,2.6E-01  &  8.6E-02  &  $\leq$\,5.0E-01  &  1.7E-01  &  0.5  & $-$ & Not detected   \\
          & SO & $v=0 \ J_{N}=6_{5}-5_{4}$ &  $\leq$\,2.6E-01  &  8.5E-02  &  $\leq$\,5.0E-01  &  1.7E-01  &  0.5  & $-$ & Not detected   \\

\hline     
\\[-2ex] 

HR\,4049  & SiO        & $v=0 \ J=1-0$     &  $\leq$\,4.5E-02  &  1.5E-02  &  $\leq$\,2.0E-01  &  6.8E-02  &  1.1  & $-$ & Not detected   \\
          &            & $v=1 \ J=1-0$     &  $\leq$\,4.2E-02  &  1.4E-02  &  $\leq$\,1.9E-01  &  6.3E-02  &  1.1  & $-$ & Not detected   \\
          &            & $v=2 \ J=1-0$     &  $\leq$\,2.6E-02  &  8.7E-03  &  $\leq$\,1.2E-01  &  3.9E-02  &  1.1  & $-$ & Not detected   \\
          & $^{29}$SiO & $v=0 \ J=1-0$     &  $\leq$\,2.7E-02  &  9.1E-03  &  $\leq$\,1.2E-01  &  4.1E-02  &  1.1  & $-$ & Not detected   \\
          & $^{30}$SiO & $v=0 \ J=1-0$     &  $\leq$\,3.0E-02  &  1.0E-02  &  $\leq$\,1.4E-01  &  4.5E-02  &  1.1  & $-$ & Not detected   \\
          & CS         & $v=0 \ J=1-0$     &  $\leq$\,9.3E-02  &  3.1E-02  &  $\leq$\,3.4E-01  &  1.1E-01  &  0.9  & $-$ & Not detected   \\
          & C$^{18}$O  & $v=0 \ J=2-1$     &  $\leq$\,2.1E-01  &  2.1E-01  &  $\leq$\,4.4E-01  &  1.5E-01  &  0.5  & $-$ & Not detected   \\
          & SO & $v=0 \ J_{N}=6_{5}-5_{4}$ &  $\leq$\,2.3E-01  &  2.3E-01  &  $\leq$\,4.6E-01  &  1.5E-01  &  0.5  & $-$ & Not detected   \\

\hline     
\\[-2ex] 

IRAS\,11472$-$0800  & SiO        & $v=0 \ J=1-0$     &  $\leq$\,3.8E-02  &  1.3E-02  &  $\leq$\,1.3E-01  &  4.3E-02  &  1.1  & $-$ & Not detected   \\
                    &            & $v=1 \ J=1-0$     &  $\leq$\,3.3E-02  &  1.1E-02  &  $\leq$\,1.1E-01  &  3.7E-02  &  1.1  & $-$ & Not detected   \\
                    &            & $v=2 \ J=1-0$     &  $\leq$\,2.9E-02  &  9.7E-03  &  $\leq$\,1.0E-01  &  3.3E-02  &  1.1  & $-$ & Not detected   \\
                    & $^{29}$SiO & $v=0 \ J=1-0$     &  $\leq$\,3.0E-02  &  9.9E-03  &  $\leq$\,1.0E-01  &  3.4E-02  &  1.1  & $-$ & Not detected   \\
                    & $^{30}$SiO & $v=0 \ J=1-0$     &  $\leq$\,2.7E-02  &  8.9E-03  &  $\leq$\,9.2E-02  &  3.1E-02  &  1.1  & $-$ & Not detected   \\
                    & CS         & $v=0 \ J=1-0$     &  $\leq$\,5.6E-02  &  1.9E-02  &  $\leq$\,1.6E-01  &  5.2E-02  &  0.9  & $-$ & Not detected   \\

\hline
\end{tabular*}

\end{center}
\small
\vspace{-1mm}
\textbf{Notes.} Uncertainties for non-detected lines are calculated based on the line width of the \doce\dosuno\ transition. Results of HD\,52961 are taken from \cite{gallardocava2022} where the complete results are reported.
\label{tab:qband}
\end{table*}

\end{document}